\documentclass{article}       
\usepackage{e}                
\begin{document}
\branch{C}   
%
\title{Electroweak corrections to the observables of $W$-boson
       production at RHIC}
\author{V. A. Zykunov\inst{1}}
\institute{
Gomel State Technical University,\\
October Av. 48, 246746, Gomel, Belarus;\\
e-mail: zykunov@ggtu.belpak.gomel.by}
\maketitle
\begin{abstract}
The processes of the single $W$-production in hadron-hadron collisions
are suggested for investigation of the nucleon spin.
An approach is proposed for the determination of
quark spin densities at low $x$.
The lowest order electroweak radiative corrections to
the observable quantities are calculated.
The numerical calculations of the cross sections
and the single spin asymmetries taking into consideration the
electroweak corrections at RHIC energies have been made.
\end{abstract}
\section{Introduction}
Despite remarkable success in solving the proton-spin problem, which was
risen by spin crisis \cite{EMC}, in experiments with fixed targets
at CERN \cite{CERN}, SLAC \cite{SLAC} and HERMES \cite{HERMES} a
conclusive solution to the problem of determination of $u-,d-,s-$ quark
and gluon polarizations has yet to be found.
Purely inclusive measurements determining the spin structure
functions for nucleons and deuteron are unfortunately
restricted to probe only certain combinations of the polarized
parton densities. A full analysis would
require additional inputs from other measurements to separate
different components.
For example, semi-inclusive measurements with SMC \cite{semi-SMC} and
HERMES \cite{semi-HERMES} allow to access to a variety of polarized
parton distributions. However, direct and separate measurements of
the polarized parton distributions will remain a limited
significance for some time.

Some new combinations of the polarized quark densities
can be obtained from the experimental data on single spin asymmetries,
when the polarized target is used in conjunction with an (un)polarized
proton beam.
This opportunity is offered by
the dedicated RHIC spin program at BNL \cite{now16} which is
supposed to start in a short period of time.
Collider polarized experiments
STAR and PHENIX at the RHIC energy ($\sqrt{S} \approx  500 GeV$)
will be the point where a high energy nucleon-nucleon
spin physics will be studied by measuring a variety of spin asymmetries.
Besides, there are some prospects on an acceleration of the polarized
protons to $1 TeV$ at Fermilab Tevatron \cite{tevat} and these will
lead to similar physics, which will be accessible to
RHIC and possibly to HERA-$\stackrel{\rightarrow}{N}$ \cite{now15}.

In this paper we are concerned with experiments which may provide
direct measurements of new independent combinations of the quark densities
in polarized nucleon. We will focus on inclusive single $W$-boson
production in hadron-hadron interactions with one longitudinally
polarized beam
\begin{equation}
N_1 \ + \stackrel{\rightarrow}{N_2} \rightarrow W^{\pm} + X \rightarrow l^{\pm} + X.
\label{1}
\end{equation}

It's well known that the single spin asymmetry of the production of
$W$-bosons via
$\ \stackrel{\ }{p}\stackrel{\rightarrow}{p} \rightarrow  W^{\pm }X$
(e.g. $A_L$ from Ref.\cite{AL})
is sensitive to the form of polarized quark distributions.
In Sec. 2 of our paper we get formulas for single asymmetries
of the process (1) for the case
when in the final state only charged lepton is detected.
Moreover, we will show the possibility of proposing reactions to study
the polarized quark densities in the region of small $x$.

Radiative events originating from the loop diagrams and
(as we consider the inclusive process)
the processes with the emission of real photons, cannot be
removed by experimental methods and so they have to be
calculated theoretically and subtracted from measured
cross sections of proposed reactions.
In this paper the total $O(\alpha)$ electroweak radiative
corrections (EWC) to the cross sections and single spin
asymmetries have been calculated.
In Sec. 3 and 4 we present the contributions of additional
virtual and real emitted particles respectively.
Numerical analysis and conclusions can be found in Sec. 5.
Some technical details of calculations are presented in Appendices.

\section{Born cross section and single asymmetries}
The cross section for the inclusive hadronic reaction (see Fig.1,1)
is given in Quark Parton Model (QPM) by conventional formula
\begin{equation}
   d\sigma_{ N_1 \stackrel{\rightarrow}{N_2} \rightarrow l^{\pm} X }=
   \int dx_1 dx_2 \sum_{i,i';r1,r2} f_i^{(1),r1}(x_1,Q^2)
                               f_{i'}^{(2),r2}(x_2,Q^2)
                   d{{{\hat \sigma}^{\pm}}}_{ii'},
\label{cs}
\end{equation}
where $f_i^{(a),r}(x,Q^2)$ is the probability of finding constituent $i$
with the fraction $x$ of the hadron's momentum and helicity $r$
(at squared momentum transfer in the partonic reaction $Q^2$)
in hadron $a$ and ${d{{\hat \sigma}^{\pm}}}_{ii'}$ is the cross section for the
elementary process leading to the final state (see Fig.1).
The summation runs over all contributing parton configurations and over
helicities of first and second partons ($r_{1,2}=\pm 1$).
The meaning of
operator "hat" will be explained below.

The process of the single $W$-boson production in hadron-hadron
collision can be described with a very good approximation
by two pair of quark-antiquark subprocesses.
So for the $W^-$ production we have
\begin{equation}
 q_i(p_1)+{\bar q}_{i'}(p_2,\eta_2) \rightarrow  W^-(q) \rightarrow
        l^{-}(k_1)+{\bar \nu}(k_2),
\label{1qp}
\end{equation}
\begin{equation}
 {\bar q}_i(p_1)+ q_{i'}(p_2,\eta_2) \rightarrow  W^-(q)\rightarrow
l^{-}(k_1)+ {\bar \nu} (k_2),
\label{2qp}
\end{equation}
and for $W^+$ one
\begin{equation}
 q_i(p_1)+{\bar q}_{i'}(p_2,\eta_2) \rightarrow  W^+(q)\rightarrow
l^{+}(k_1)+ \nu (k_2),
\label{3qp}
\end{equation}
\begin{equation}
 {\bar q}_i(p_1)+ q_{i'}(p_2,\eta_2) \rightarrow  W^+(q)\rightarrow
l^{+}(k_1)+ \nu (k_2).
\label{4qp}
\end{equation}
Our notations are the following (see Fig.1,2):
$p_1$ is a 4-momentum of a first unpolarized (anti)quark with flavor $i$
and mass $m_1$;\
$p_2$ is a 4-momentum of a second (anti)quark with flavor $i'$ ($i'$ denotes
the weak isospin partner of the quark $i$), mass $m_2$
and polarization vector $\eta_2$;\
$k_1$ is a 4-momentum of a final charged lepton $l^-$ or $l^+$ with mass $m_l$;\
$k_2$ is a 4-momentum of (anti)neutrino,
$q=p_1+p_2$ is a 4-momentum of the $W$-boson with mass $m_W$.
We use the standard set of Mandelstam invariants for the partonic elastic
scattering
\begin{equation}
s=(p_1+p_2)^2,\ t=(p_1-k_1)^2,\ u=(k_1-p_2)^2.
\end{equation}

Squaring the matrix elements of partonic subprocess
and (we use the covariant expression for
the polarization vector of $i'$ parton from \cite{vec_pol})
we get the invariant parton-parton cross section in the Breit-Wigner form
\begin{equation}
 d \sigma_{ii'}^{\pm} = \frac{\alpha^2}{4N_cs_w^4}
            \frac{|V_{ii'}|^2 B_{ii'}}
            {s((s-m_W^2)^2+m_W^2\Gamma_W^2)}
            \delta (p_1+p_2-k_1-k_2)
            \frac{d^3k_1}{{k_1}_0}
            \frac{d^3k_2}{2{k_2}_0},
\end{equation}
where \\
$1/N_c=1/3$ is a color factor,
$s_w=\sqrt{1-c_w^2}$ is a sine of the weak mixing angle,
$c_w=m_W/m_Z,$
$m_Z$ is the $Z$-boson mass,
$\Gamma_W$ is the $W$-boson width,
$V_{ii'}$ is Cabibbo-Kobayashi-Maskawa mixing matrix,\\
${\displaystyle \hbox to 0.5cm{\hfill}
  B_{ii'}= \left\{
        \begin{array}{cc}
u^2(1-r_2p_{N_2}) \mbox{ \ \ \ \ \ \ for (\ref{1qp}) subprocess,  }\\
t^2(1-r_2p_{N_2}) \mbox{ \ \ \ \ \ \ for (\ref{2qp}) subprocess,  }\\
t^2(1+r_2p_{N_2}) \mbox{ \ \ \ \ \ \ for (\ref{3qp}) subprocess,  }\\
u^2(1+r_2p_{N_2}) \mbox{ \ \ \ \ \ \ for (\ref{4qp}) subprocess,  }
        \end{array}
        \right.
}$ \hfill \\
and $p_{N_2}$ is the degree of longitudinal polarization of a
second hadron such that $p_{N_2}=\pm 1$.

The integration w.r.t. 4-momenta of unobservable (anti)neutrino gives
\begin{equation}
\int  \frac{d^3k_2}{2{k_2}_0}
        \delta (p_1+p_2-k_1-k_2) =
        \delta (s+t+u-m_1^2-m_2^2-m_l^2).
\end{equation}
Then in accordance with QPM we substitute $p_{1(2)} \rightarrow x_{1(2)}P_{1(2)}$,
where $P_{1(2)}$ is 4-momenta of initial nucleons with masses $m_N$,
$x_{1(2)}$ is the fraction of the first (second) nucleon momentum
that is carried by the incoming quarks.
We shall denote this procedure by operator "hat".
Then we multiply by the parton densities of first and second hadrons,
sum over helicities of quarks and integrate over $x_1$, $x_2$
according to formula (\ref{cs}).

Let us introduce the Mandelstam variables for hadronic reaction
\begin{equation}
S=2P_1P_2,\ T=-2P_1k_1,\ U=-2P_2k_1,
\end{equation}
(then $\hat s=x_1x_2S+m_1^2+m_2^2,\ \hat t=x_1T+m_1m_N+m_l^2,\
  \hat u=x_2U+m_2m_N+m_l^2$)
and integrate w.r.t. $x_2$
with help of $\delta$-function and noting that in QPM
$$   \delta (\hat s + \hat t + \hat u -m_1^2-m_2^2-m_l^2) =
   \frac{1}{D} \delta (x_2 + \frac{x_1(T+m_N^2)+m_l^2}{D}),
   \ D=x_1S+U+m_N^2.$$
We can see that in this case $x_2=x_2^0 \equiv -(x_1(T+m_N^2)+m_l^2)/D $
and this substitution
corresponds to Born kinematics and we denote this by subscript
(or superscript) "0".

Finally, let us consider the general form of the cross section
of hadronic process (\ref{1}).
In the hadron-hadron collisions the center of parton-parton masses frame
has an undetermined motion along the beam direction. Therefore
we use the standard in this case variables:
centre-of-mass energy ($\sqrt{S}$), component of the
detected particle vector transverse to the beam direction
($|{k_1}_{\perp}| \equiv {k_1}_T$),
and pseudorapidity ($\eta$), which approximately equals rapidity $y$,
since in our case $m_l \ll {k_1}_T$; then we have for $T$ and $U$
\begin{equation}
 T=-\sqrt{S}{k_1}_Te^{-\eta},\
   U=-\sqrt{S}{k_1}_Te^{\eta}.
\end{equation}
Integrating w.r.t. azimuth $\Phi$ (it's possible
as the first initial hadron is unpolarized and the second one is
longitudinally polarized) we have phase space
$ d^3k_1/{k_1}_0 \Rightarrow \pi d\eta {k_1}_T^2 $,
and hence the Born cross section has the form
\begin{equation}
 \sigma^{\pm}_0
 = \sum_{i,i'} \int dx_1 q_i(x_1,Q^2) \Sigma_0,
\label{CS_born}
\end{equation}
where $\sigma^{\pm}$ denotes the double differential cross section
\begin{equation}
  \sigma^{\pm}
  \equiv \frac{ d \sigma_{ N_1 \stackrel{\rightarrow}{N_2} \rightarrow l^{\pm} X} }
              {d\eta d{k_1}_T^2}.
\label{abbr}
\end{equation}
We use this abbreviation as well as the following
\begin{equation}
 \sigma^{\pm}  = \bar \sigma^{\pm} + p_{N_2} \Delta \sigma^{\pm},
\end{equation}
where  $\bar \sigma^{\pm} ( \Delta \sigma^{\pm} )$ is the spin averaged
(polarization) part of cross section in the whole text.

The Born cross section is proportional to factor
$  \Sigma_0 = \Sigma (x_2^0)$,\\
where
\begin{equation}
  \Sigma (x_2)
  = \frac{ \pi \alpha^2}{4N_cs_w^4}
    \frac{ |V_{ii'}|^2 {\hat B}_{ii'}  }
         {{\hat s}((\hat s-m_W^2)^2+m_W^2\Gamma_W^2)D }
    F_{i'}^{(2)}(x_2,Q^2),
\label{Sigma}
\end{equation}
In the expressions (\ref{CS_born}),(\ref{Sigma})
the combination of the quark densities for a second nucleon
has the form
\begin{equation}
F_{i'}^{(2)}(x_2,Q^2)=\bar q_{i'}(x_2,Q^2)
                      -c_{i'}p_{N_2}\Delta q_{i'}(x_2,Q^2),
\end{equation}
where $(\Delta)q_i(x,Q^2)=f_i^{+}(x,Q^2) (-)+ f_i^{-}(x,Q^2)$
are (longitudinally polarized) spin averaged quark densities,
$c_{i'}=-1(+1)$ for quark(antiquark).

The form of the Born cross section (\ref{CS_born}) is convenient as
the factorized part in the so-called 'soft' $O(\alpha^3)$ cross section
(see Sect.3,4) and here let us present more simple expression for
the Born cross section of the processes (\ref{1}) as a convolution of
partonic cross section with two parton distribution functions which
we will use in the rest of this section
\begin{equation}
\sigma^{\pm}_0
  = \frac{ \pi \alpha^2}{4N_cs_w^4}
    \sum_{i,i'} \int dx_1
    \frac{ |V_{ii'}|^2 {\hat B}_{ii'}^0  }
         {{\hat s_0}((\hat s_0-m_W^2)^2+m_W^2\Gamma_W^2)D }
    q_i(x_1,Q^2) F_{i'}^{(2)}(x_2^0,Q^2).
\label{CS_born2}
\end{equation}

Let us define the polarization single asymmetries as
normalized difference of the cross sections with specific spin configurations
\begin{equation}
  A^{l^{\pm}}(\eta,{k_1}_T)=
\frac{\sigma^{\pm}(p_{N_2}=1)-\sigma^{\pm}(p_{N_2}=-1)}
     {\sigma^{\pm}(p_{N_2}=1)+\sigma^{\pm}(p_{N_2}=-1)} =
\frac{\Delta \sigma^{\pm}}{\bar \sigma^{\pm}}.
\end{equation}
Supposing that CKM matrix has diagonal form  $V_{ii'}=\delta_{ii'}$
we can see that flavors of initial quarks and antiquarks are
$ i= d, s, b;\ i'=\bar u, \bar c, \bar t \ $ for (\ref{1qp}) subprocess,
$ i=\bar u, \bar c, \bar t;\ i'= d, s, b \ $ for (\ref{2qp}) subprocess,
$ i=u, c, t;\ i'=\bar d, \bar s, \bar b \  $ for (\ref{3qp}) subprocess,
$ i=\bar d, \bar s, \bar b;\ i'= u, c, t \ $ for (\ref{4qp}) subprocess.
Neglecting the contributions of the heavy quarks $(c,b,t)$
(the strange quark contribution $s \bar u \rightarrow W^-$
is neglected too as we suppose $V_{s\bar u} $ equals zero)
we can write the Born single asymmetries as
\begin{equation}
  A^{l^+}_0(\eta,{k_1}_T)=
-\frac{\int dx_1(u'(x_1)\Delta \bar d(x_2^0)-\bar d'(x_1)\Delta u(x_2^0))}
      {\int dx_1(u'(x_1) \bar d(x_2^0)+\bar d'(x_1) u(x_2^0))},\
\label{a1}
\end{equation}
\begin{equation}
  A^{l^-}_0(\eta,{k_1}_T)=A^{l^+}_0(\eta,{k_1}_T)(u \leftrightarrow d).
\label{a2}
\end{equation}
Here
$$ u'(x_1)=K_t u(x_1),\ \bar u'(x_1)=K_t \bar u(x_1),\
  d'(x_1)=K_u d(x_1),\ \bar d'(x_1)=K_u \bar d(x_1), $$
\begin{equation}
 K_{t(u)}=\frac{{\hat t}^2({\hat u}^2)}
{{\hat s}((\hat s-m_W^2)^2+m_W^2\Gamma_W^2)D }|_{x_2=x_2^0}
\label{K}
\end{equation}
(we suppress the argument $Q^2$ in the quark distributions of formulas
(\ref{a1},\ref{K})).

The physically allowed region of $x_1$ and $x_2$  (see Fig 2.)
is given by
\begin{equation}
 -\frac{U+m_N^2-m_l^2}{S+T+m_N^2} \leq x_1 \leq 1, \ \ \
                      x_2^0 \leq x_2 \leq 1.
\end{equation}
Let us remark that in the region of large $x_1$ and small
${k_1}_T/\sqrt{S}$ the expression
$x_2^0$ does almoust not depend on $x_1$ ($x_2^0 \approx -T/S$).
Dividing the region of integration in (\ref{a1},\ref{a2})
by parameter $x_1^*$
(we can choose such a value of $x_1^*$ in order to have well defined
polarized quark densities in the region  $x_1 < x_1^*$)
we obtain the expression
\begin{equation}
\Delta u(-T/S)      \int_{x_1^*}^1 dx_1 \bar d'(x_1)\ - \
\Delta \bar d(-T/S) \int_{x_1^*}^1 dx_1  u'(x_1) =
\label{final}
\end{equation}
$$ =A^{l^+}_0(\eta,{k_1}_T)
   \int_{x_1^{min}}^{1} dx_1(u'(x_1) \bar d(x_2^0)+\bar d'(x_1) u(x_2^0))$$
$$ - \int_{x_1^{min}}^{x_1^*} dx_1
   (\Delta u(x_2^0)\bar d'(x_1) -\Delta \bar d(x_2^0) u'(x_1) ) $$
and analogously for $A^{l^-}$ by replacing $u \leftrightarrow d$.

Off-shell approach, which we used gives the possibility
to investigate the polarized quark densities in the region
of very small $x$.
Really, the expression $-T/S$ under conditions of collider experiment
can reach
very small values, so using, for example, RHIC kinematics point
$\sqrt{S}=500 GeV$,
$ {k_1}_T = 10GeV$, $\eta=2$ we can see that  $-T/S$ does not
exceed 0.0027.
That minimum value of $x$ is about one order of magnitude
below the values quoted in studies where
on-shell approximation for the $W$-boson was used.

So, equations (\ref{final}) connect the polarized quark densities
in the region of small $x$ with the observable single asymmetries,
combination of unpolarized quark densities
and polarized quark densities in the region where they are
well defined.
If the three of the supplementary measurable quantities (for example
double asymmetries $A_{\pm}(x,y)$ from Ref.\cite{my3} and QPM-expression
for $g_1(x)$) are used, equations (\ref{final}) allow to determine
the low-$x$ behavior of polarized $u-$ and $d-$quark and antiquark
distributions in nucleon.

\section{Contribution of additional virtual particles to the cross section}
To extract the reliable data on single asymmetries with high precision
from hadron collider
experiment, it is necessary to consider higher order electroweak
radiative corrections.
The final state photonic corrections to
processes of $W$-production in unpolarized pp-collisions were calculated
in Ref.\cite{FSR}.
More accurate calculation of these electroweak corrections
have been suggested in Ref.\cite{pp}, where both initial and final state
radiation have been included, and the collinear singularity
associated with final state photon radiation is regularized by the lepton
mass. On the contrary, the collinear singularities associated with initial
state radiation are subtracted and adsorbed into the parton distribution
functions. Due to these results do not contain the quark masses, but
depend analytically on other unphysical parameter $\delta_{\theta}$,
which determines a collinear region.

In this paper we present the new explicit
formulae for EWC to inclusive single $W$-production in polarized hadron-hadron
collisions, where by analogy to the calculations of the radiative
corrections to the hadron current in the deep inelastic scattering
of lepton by nucleons \cite{bardin,bom_z,BS2,my1} we used a finite quark
masses to regularise the collinear singularities. In Sect.5 we consider
the dependence of the results on various quark masses and
the comparison of them with the results for EWC, when the strategy
of collinear singularity extraction has been employed.

As in final state only charged lepton is detected
we use for calculation
the covariant method \cite{covar}, which has conclusive advantage:
the results are independent of unphysical
parameter -- maximum soft photon energy $E_{cut}$
(see for example Ref.\cite{pp}).

The one-loop contribution of additional virtual particles
(V-contribution) has been calculated in t'Hooft-Feynman gauge
and in on-mass renormalization scheme which uses
$\alpha, m_W, m_Z,$ Higgs boson mass $m_H$ and the fermion masses
as independent parameters. The virtual one-loop diagrams are shown
in Fig.3.

The cross section of V-contribution is proportional to the Born cross
section and could be written as
\begin{equation}
 \sigma _V^{\pm}
 = \sum_{i,i'} \int dx_1 q_i(x_1,Q^2)
   \hat \delta^{ii'} _{V}|_{x_2=x^0_2}
   \Sigma_0.
\label{v}
\end{equation}
Here factor $\delta^{ii'} _{V} $ is polarization independent and
consists of seven terms
\begin{equation}
\delta^{ii'} _V = \delta _W +\delta _{Vl} +\delta^{ii'}_{Vq} +
\delta_{Sl}+ \delta_{Sq} +\delta^{ii'}_{\gamma W}+\delta^{ii'}_{ZW}.
\label{delta}
\end{equation}
We do not repeat here the explicit expressions for all these
contributions but instead refer the readers to Ref.\cite{BS2},\cite{BS1}.
Let us consider the meaning of various terms in (\ref{delta}).
The correction $\delta _W$ is the
$W$-boson self-energy contribution (Fig.3,1), which has a slightly
modified form from what is presented in Ref.\cite{BS2}. This is due to
resonant $W$ boson production
\begin{equation}
\delta _W = 2 Re \frac{s-m_W^2-im_W\Gamma_W}{(s-m_W^2)^2+m_W^2\Gamma_W^2}
            {\hat \Sigma}^W_T(s) ,
\end{equation}
where
${\hat \Sigma}^W_T(s)$ is the renormalized transverse part of the
$W$-boson self energy
(it can be found from formula (6.1) of Ref.\cite{BS1}).
The term $\delta _{Vl}$ is the leptonic vertex correction (Fig.3,2)
(see formula (2.10) of Ref.\cite{BS2}),
$\delta^{ii'} _{Vq}$ is the quark vertex correction (Fig.3,3)
(see formula (2.11) of Ref.\cite{BS2}, we left the flavour indices at
this correction and boxes contributions),
$\delta _{Sl}$ is the neutrino self energy contribution (Fig.3,4)
(see formula (2.12) of Ref.\cite{BS2}).
Up-quarks self energy contribution (Fig.3,5) (down-quarks
self energy contribution equals zero) is given by
\begin{equation}
\delta _{Sq}
= -\frac {\alpha}{4\pi} \biggl[Q^2 _u \left ( \ln \frac {m_Z^2}
{m^2_u} -2 \ln \frac {m^2_u} {\lambda ^2} \right )
-Q^2_d \left( \ln \frac {m_Z^2} {m^2_d}
-2 \ln \frac {m^2 _d} {\lambda ^2}
\right) +\frac{3}{2} \biggr].
\end{equation}
This expression is obtained from formula (5.46) of Ref.\cite{BS1}
(see also the remark below it).

We recalculated the box contributions through the
quantities $I^{\gamma W}_{1,2}$ and $I^{ZW}_{1,2}$ of Ref.\cite{BS2}
and present them here to stress their dependence on flavour of
quarks and the channel of reaction:
$\gamma W$ box contribution (Fig.3,6 and Fig.3,7)
\begin{equation}
\delta^{u \bar d} _{\gamma W} =
 \delta^{\bar u d} _{\gamma W} =
 \delta _{\gamma W},\ \
 \delta^{d \bar u} _{\gamma W} =
 \delta^{\bar d u} _{\gamma W} =
 \delta _{\gamma W}(t \leftrightarrow u, i \leftrightarrow i'),
\end{equation}
where
\begin{equation}
\delta _{\gamma W} =\frac {2\alpha}{\pi} Re Q_l
(Q_iI^{\gamma W, i}_1(s,t)+Q_{i'}I^{\gamma W, i'}_2(s,u));
\end{equation}
$ZW$ box contribution (Fig.3,6 and Fig.3,7)
\begin{equation}
\delta^{u \bar d} _{ZW} =
 \delta^{\bar u d} _{ZW} =
 \delta _{ZW},\ \
 \delta^{d \bar u} _{ZW} =
 \delta^{\bar d u} _{ZW} =
 \delta _{ZW}(t \leftrightarrow u, i \leftrightarrow i'),
\end{equation}
where
\begin{eqnarray}
\delta _{ZW} &=&\frac {2\alpha}{\pi} Re [
((v_l+a_l)(v_i+a_i)
+(v_{\nu}+a_{\nu})(v_{i'}+a_{i'}))I_1^{ZW}(s,t)
\nonumber \\&&
+((v_l+a_l)(v_{i'}+a_{i'})
+(v_{\nu}+a_{\nu})(v_i+a_i))I_2^{ZW}(s,u)].
\end{eqnarray}

The vector and axial vector couplings constants $v_i,\ a_i$ of the
$i$-fermion--$Z$ vertex are determined from
the electric charge ($Q_i$) of fermion expressed in units of
proton's charge (e.g. $Q_u \equiv Q_{\bar u}=+2/3$,
$Q_l=-1$) and
the 3-component of the weak fermion isospin ($I_i^3$)
$$ v_i=\frac{I_i^3-2s_wQ_i}{2s_wc_w},\ \
   a_i=\frac{I_i^3}{2s_wc_w}.$$

Let us present the cross section of V-contribution as the sum of
infrared divergent (IR) and IR-finite parts
\begin{equation}
  \sigma^{\pm}_V=\sigma^{\pm,IR}_V+\sigma^{\pm,F}_V
= \sum_{i,i'} \int dx_1 q_i(x_1,Q^2)
  \Sigma_0
  ( \hat \delta^{IR}_{V} + \hat \delta^{F}_{V} )|_{x_2=x^0_2}.
\label{IRV}
\end{equation}
Infrared divergent part of V-contribution
is regularized with the help of photon mass $\lambda$ and for
the correction $\delta^{IR}_{V}$  we find
\begin{equation}
\delta^{IR}_V=\frac {\alpha}{2\pi}\log \frac s {\lambda ^2}J(0),
\end{equation}
where the expression for $J(0)$ will be considered in the next section.
The finite part of V-contribution cross section contains correction
\begin{equation}
\delta^{F}_V=
\delta^{ii'}_V - \delta^{IR}_V
 =\delta^{ii'}_V(\lambda^2 \rightarrow s).
\end{equation}

\section{ The photon bremsstrahlung
$ N_1 \stackrel{\rightarrow}{N_2} \rightarrow l^{\pm} \gamma X $}
In order to get infrared finite results for the hadronic process
cross section we have to
include the real bremsstrahlung correction (Fig.4).

The differential cross section for the partonic process with emission
of one real photon reads
\begin{equation}
 d\sigma_{q_i \stackrel{\rightarrow}{q_{i'}} \rightarrow l\nu \gamma}
= \frac{\alpha^3|V_{ii'}|^2}{2^6\pi^2s_w^4N_c}\frac{1}{s}
   \sum_{pol}^{-} |R|^2  \delta (p_1+p_2-k_1-k_2-k)
            \frac{d^3k_1}{2{k_1}_0}
            \frac{d^3k_2}{2{k_2}_0}
            \frac{d^3k}{2{k}_0},
\end{equation}
where squared matrix element\footnote{
For the calculation of matrix element we used the standard
Feynman rules and the procedure of separation of diagram
Fig.4,4  into the initial and final state radiation parts according to
Ref.\cite{FSR}. As a result, 'initial' part of our matrix element
completely coincides with formula (2.14) of Ref.\cite{FSR}.}
is the sum of initial state (index $i$),
final state (index $f$) and interference terms respectively
\begin{equation}
 \sum_{pol}^{-} |R|^2 =
(R_{i}R_{i}^+)^{nn'} + (R_{f}R_{f}^+)^{nn'} +
(R_{i}R_{f}^+ +R_{f}R_{i}^+)^{nn'},
\end{equation}

$$
(R_{i}R_{i}^+)^{nn'} =
-\Pi_q\Pi_q^+
Sp[ {(G_i^{nn'})}^{\mu \rho} U_{i,p}^n
    {({G_i^{nn'}}^T)}^{\mu '}_{\rho} U_{i,a}^n ]
Sp[ \gamma_{\mu} U_{f,a}^{n'} \gamma_{\mu '} U_{f,p}^{n'} ],
$$
$$
(R_{f}R_{f}^+)^{nn'} =
-\Pi_l\Pi_l^+
Sp[ \gamma_{\mu} U_{i,p}^{n} \gamma_{\mu '} U_{i,a}^{n} ]
Sp[ {(G_f^{n'})}^{\mu \rho} U_{f,a}^{n'}
    {({G_f^{n'}}^T)}^{\mu '}_{\rho} U_{f,p}^{n'} ],
$$
$$ (R_{i}R_{f}^+ + R_{f}R_{i}^+ )^{nn'} =
 - \Pi_l\Pi_q^+
(\ Sp[ {(G_i^{nn'})}^{\mu \rho} U_{i,p}^n \gamma_{\mu '} U_{i,a}^n ]
Sp[ \gamma_{\mu} U_{f,a}^{n'} {{(G_f^{n'}}^T)}^{\mu '}_{\rho} U_{f,p}^{n'} ]
$$
$$ + Sp[ \gamma_{\mu} U_{i,p}^n {({G_i^{nn'}}^T)}^{\mu '}_{\rho} U_{i,a}^n ]
Sp[ {(G_f^{n'})}^{\mu \rho} U_{f,a}^{n'} \gamma_{\mu '} U_{f,p}^{n'} ] \ ).
$$
Indices $n,\ n'$ denote the type of the processes:
$n=-(+)$ for $q \bar q (\bar q q)$ initial state, and
$n'=-(+)$ for $l^- \bar \nu (l^+ \nu)$ final state, and
$\Pi_l (\Pi_q) $ is a $W$-boson propagator in the case of the lepton(quark)
bremsstrahlung
\begin{equation}
 \Pi_l=1/(s-m_W^2+im_W\Gamma _W),
 \ \Pi_q=-1/(s-2kq-m_W^2+im_W\Gamma _W)=1/(z+t_{w\Gamma}),
\end{equation}
$$ t_{w\Gamma}=t_w-im_W\Gamma _W, \ \ t_w=v-s+m_W^2, q=p_1+p_2.\ $$

As the kinematical variables of the radiated process we use in this case
\begin{eqnarray}
 z=2kk_1, \; z_1=2kp_1,\; t_1=(p_2-k_2)^2,\; u_1=2kp_2=v+z-z_1,
\nonumber \\
 v=2kk_2=s+u+t-m_1^2-m_2^2-m_l^2,
\end{eqnarray}
where $k$ is a 4-momentum of the radiated photon.

The matrices $U$ originate from the products of bispinor amplitudes and
the expressions $(1 \pm \gamma_5)$:
\begin{equation}
 U_{i,p}^- = (1-\gamma_5)\hat p_1,\
   U_{i,a}^- = (1+\gamma_5 \hat \eta_2)(\hat p_2 - m_2),\
   U_{i,p}^+ = (1-\gamma_5)\hat p_2,\
   U_{i,a}^+ = \hat p_1 - m_1,
\end{equation}
$$ U_{f,a}^- = (1-\gamma_5)\hat k_2,\
   U_{f,p}^- = \hat k_1 + m,\
   U_{f,a}^+ = (1-\gamma_5)\hat k_1,\
   U_{f,p}^+ = \hat k_2,\ (\hat p=\gamma^{\mu}p_{\mu})$$
and the
matrices $G$ originate from the fermion propagator and $WW\gamma$-vertex:
\begin{equation}
 {(G_i^{-n'})}^{\mu \rho} =
   Q_i\gamma^{\mu}\frac{2p_1^{\rho}-\hat k \gamma^{\rho}}{z_1} -
   Q_{i'}\frac{2p_2^{\rho}-\gamma^{\rho}\hat k}{u_1}\gamma^{\mu} +
   {(G_i^{W^{n'}})}^{\mu \rho},
\end{equation}
$$ {(G_i^{+n'})}^{\mu \rho} =
   Q_{i}\frac{-2p_1^{\rho}+\gamma^{\rho}\hat k}{z_1}\gamma^{\mu} -
   Q_{i'}\gamma^{\mu}\frac{-2p_2^{\rho}+\hat k \gamma^{\rho}}{u_1} +
   {(G_i^{W^{n'}})}^{\mu \rho}, $$
$$ {(G_f^{-})}^{\mu \rho} =
   Q_{l}\frac{2k_1^{\rho}+\gamma^{\rho}\hat k}{z}\gamma^{\mu} +
   {(G_f^{W^-})}^{\mu \rho},\
   {(G_f^{+})}^{\mu \rho} =
   -Q_{l}\gamma^{\mu}\frac{2k_1^{\rho}+\hat k \gamma^{\rho}}{z} +
   {(G_f^{W^+})}^{\mu \rho}, $$
$$ {(G_f^{W^+})}_{\mu \rho} =
   \gamma^{\rho '}\frac{1}{2kq}C^3_{\rho \mu \rho '}(-k,q,k-q),\
  {(G_f^{W^-})}_{\mu \rho} =
   \gamma^{\nu '}\frac{1}{2kq}C^3_{\rho \nu '\mu}(-k,k-q,q), $$
$$ {(G_i^{W^+})}_{\mu \rho} =
   \gamma^{\nu '}\frac{1}{2kq}C^3_{\rho \nu '\mu} (-k,q,k-q),\
  {(G_i^{W^-})}_{\mu \rho} =
   \gamma^{\rho '}\frac{1}{2kq}C^3_{\rho \mu \rho '}(-k,k-q,q), $$
where the matrix $C^3$ corresponds to $WW\gamma$-vertex and for
the minimal standard model reads
$$ C^3_{\mu \nu \rho} (p^0,p^+,p^-) =
     g_{\rho \nu} (p^- - p^+)_{\mu}
     + g_{\mu \nu} (p^+ - p^0)_{\rho}
     + g_{\mu \rho } (p^0 - p^-)_{\nu}. $$

After transition to hadron-hadron cross section according to
prescription of QPM,
we can present the cross section of bremsstrahlung
(R-contribution) by splitting it into a soft IR-part and a hard
contribution \cite{covar} (we again use the abbreviation (\ref{abbr})
for the double cross section $\sigma^{\pm}$)
\begin{equation}
\sigma_R^{\pm}=
    \sigma_R^{\pm,IR}
  + \sigma_R^{\pm,F} .
\label{sir}
\end{equation}
The infrared divergent part (the first term) of expression (\ref{sir}) reads
(we made the substitution $dx_2=d\hat v/D$)
\begin{equation}
  \sigma_R^{\pm,IR} =
\sum_{i,i'} \int dx_1 q_i(x_1,Q^2) {\hat \Sigma}_R^{IR},
\label{IRR}
\end{equation}
where
\begin{equation}
\hat \Sigma_R^{IR}
= -\frac{\alpha}{\pi} \int_{\hat v_{min}}^{\hat v_{max}} d\hat v
{\Sigma(x_2)} I[\hat F^{IR}].
\end{equation}
The procedure of integration over the photon phase space
defined as $I[A]$ is described by (\ref {int}) in Appendix A.
And for $ F^{IR}$,
using the method of separation of IR-terms \cite{covar} we find
\begin{equation}
 F^{IR}=Q_l^2\frac{m_l^2}{z^2}
         +c_lQ_lQ_i\frac{t}{zz_1}
         -c_lQ_lQ_{i'}\frac{u}{zu_1}
         +Q_i^2\frac{m_1^2}{z_1^2}
         -Q_iQ_{i'}\frac{s}{z_1u_1}
         +Q_{i'}^2\frac{m_2^2}{u_1^2},
\end{equation}
where
 \begin{equation}
  c_l= \biggl \{
         \begin{array}{cc}
 +1 \ \ \  \mbox{ for (\ref{1qp}) and (\ref{4qp})  subprocesses,  }\\
 -1 \ \ \  \mbox{ for (\ref{2qp}) and (\ref{3qp})  subprocesses. }
         \end{array}
 \end{equation}

Introducing
$J(\hat v)=\hat v \lim \limits_{\lambda \to 0} I[\hat F^{IR}]$
we obtain two parts of the soft cross section:
\begin{equation}
 \hat \Sigma_R^{IR} =
     -\frac{\alpha}{\pi} {\Sigma}_0 J(0)
      \int \limits _{\hat v_{min}}^{\hat v_{max}}\frac{d\hat v}{\hat v}
          +\frac{\alpha}{\pi}
     \int \limits _{\hat v_{min}}^{\hat v_{max}} d\hat v
          (\Sigma _0 J(0)/\hat v - {\Sigma(x_2)} I[\hat F^{IR}]).
\end{equation}
The second term is infrared free and lower limit of integration for it
equals zero, and the first one contains infrared divergence.
In the center-of-mass-system of initial hadrons the limits of
integration are
\begin{eqnarray}
\hat v_{min}=(\stackrel{\rightarrow}{k} =0)=2\lambda {k_{2}}_0 = \lambda \tilde v,\ \
\tilde v \approx  \frac{D^2+TU}{D\sqrt{S}},
\nonumber \\
\hat v_{max}=({x_2}_{max}=1)=D(1-x_2^0)
\end{eqnarray}
and the infrared divergent part of bremsstrahlung cross section
has the form
$ -{\alpha}/{\pi} \Sigma_0 J(0)\log ({\hat v_{max}}/{\lambda \tilde v })$.

Summing up IR-parts of the V- and R- contributions (formulas (\ref{IRV})
and (\ref{IRR})) we get
\begin{eqnarray}
& \sigma_R^{\pm,IR}&
+ \sigma_V^{\pm,IR} =
   \sum_{i,i'} \int dx_1 q_i(x_1,Q^2)
   \frac{\alpha}{2\pi}\Sigma_0  J(0)
  \log \frac{ {\tilde v}^2 {\hat s}_0 }{\hat v_{max}^2}
\nonumber \\
&&
 +  \sum_{i,i'} \int dx_1 q_i(x_1,Q^2)
   \frac{\alpha}{\pi}
   \int \limits _{0}^{\hat v_{max}} d\hat v
\frac{\Sigma _0 J(0)-\Sigma(x_2) J(\hat v)}{\hat v},
\end{eqnarray}
as a result infrared divergences completely cancel out.

During the calculation of $I[\hat F^{IR}]$
the following expressions are used
\begin{eqnarray}
I[\frac{1}{z^2}]=\frac{1}{m_l^2v},
&&I[\frac{1}{z_1^2}]=\frac{1}{m_1^2v},
\nonumber \\
I[\frac{1}{zz_1}]=-\frac{1}{vt}\log\frac{t^2}{m^2_lm_1^2},
&&I[\frac{1}{zu_1}]=-\frac{1}{vu}\log\frac{u^2}{m_l^2m_2^2},
\nonumber \\
I[\frac{1}{z_1u_1}]=\frac{1}{vs}\log\frac{s^2}{m_1^2m_2^2},
&&I[\frac{1}{u_1^2}]=\frac{1}{m_2^2v},
\end{eqnarray}
and therefore $ J(\hat v) $ has the form (we remind
that $\hat v=D(x_2-x_2^0)$)
\begin{eqnarray}
\label{j}
 J(\hat v)&=&
 Q_l^2 - c_lQ_lQ_i \log\frac{\hat t^2}{m^2_lm_1^2} +
 c_lQ_lQ_{i'} \log\frac{\hat u^2}{m_l^2m_2^2}
\nonumber \\
 &&
 + Q_i^2 - Q_iQ_{i'}\log\frac{\hat s^2}{m_1^2m_2^2}  + Q_{i'}^2.
\end{eqnarray}

After extraction of infrared singularity from partonic process cross section
and integrating over whole phase space of real photon (see Appendix A)
we can present the
hard contribution cross section of inclusive process
$ N_1 \stackrel{\rightarrow}{N_2} \rightarrow l^{\pm} \gamma X $ in the form
\begin{equation}
 \sigma^{\pm,F}_R =
   \sum_{i,i'} \int dx_1dx_2 q_i(x_1,Q^2) F_{i'}^{(2)}(x_2,Q^2)
 {\hat \Sigma}^F_R,
\end{equation}
where
\begin{eqnarray}
\Sigma^F_R  &=&\frac{\alpha^3}{8s_w^4sN_c}|V_{ii'}|^2
    (Q_l^2\Pi_l\Pi_l^+V_l+Q_l Re[\Pi_lV_{lq}]+V_q
\nonumber \\
   &&+Q_l\Pi _l\Pi_l^+ Re[V_{lw}]+ Re[\Pi_l]V_{qw}+\Pi_l\Pi_l^+V_w),
\end{eqnarray}
and
$$\Pi_l\Pi_l^+= 1/((s-m_W^2)^2+m_W^2\Gamma_W^2).$$
The subscript in $V$ means the origin of bremsstrahlung photon:
$l/q/w$ refers to the radiation from lepton (Fig.4,3) /
quark (Fig.4,1 and Fig.4,2) / $W$-boson (Fig.4,4) legs respectively.
Double subscript corresponds to the same interference term.
The expressions for the quantities $V$ can be found in Appendix B.

So, we get three terms of $O(\alpha^3)$ cross section which
include radiative corrections:
infrared finite part of V(and R)-contribution $\sigma^{\pm,F}_{V(R)}$
and the part $(\sigma^{\pm,IR}_V + \sigma^{\pm,IR}_R)$, where
infrared divergence cancelled out in the sum
\begin{equation}
\sigma^{\pm}_{EWC} =  \sigma^{\pm,F}_{V} + \sigma^{\pm,F}_{R}
        + (\sigma^{\pm,IR}_V + \sigma^{\pm,IR}_R).
\end{equation}
All of these terms depend on a second hadron
polarization. This dependence is factorized in the expression
$ F_{i'}^{(2)}(x_2,Q^2) $, which is containing only in
$\Sigma(x_2)$ and $\Sigma_0$. Such a factorization gives the possibility
to obtain the cross section of a single $W$-production in unpolarized
nucleon-nucleon collisions by doing a simple replacment
\begin{equation}
 F_{i'}^{(2)}(x_2,Q^2) \rightarrow q_{i'}(x_2,Q^2).
\end{equation}

\section{Discussion of numerical results and conclusions}
In the following the scale of radiative corrections and their effect
on the observables of the processes (1)
will be discussed
\footnote{A program of numerical calculations of the cross sections
and single spin asymmetries
of processes (1) (in FORTRAN) including $O(\alpha)$
electroweak corrections is available by contacting to author via e-mail}.
First of all we compare the numerical estimations of our formulas
to the results of the calculations Ref.\cite{pp} for the unpolarized
$p \bar p \rightarrow l^{\pm} X$.
We have calculated (Fig.5) $\mu^+$ transverse momentum spectrum
as a function of ${k_1}_T$ for the process
$p \bar p \rightarrow \mu^{+} X$ at $\sqrt{S}=1.8 TeV$ (Tevatron kinematics).
The integration over pseudorapidity ($-1.2 \leq \eta \leq 1.2$) had been done
and the MRS LO 98 set of unpolarized parton distribution functions
\cite{MRS98} are used.
It can be seen from comparison of Fig. 5 and Fig.7,b of Ref.\cite{pp}
the total EWC have a similar behavior: $O(\alpha^3)$ cross section
in the region $25GeV<{k_1}_T<34 GeV$ is a little bit higher
than Born cross section; the EWC contribution to the cross section
is insignificant at ${k_1}_T \sim 35(37)GeV$ for Ref.\cite{pp}(our)
calculations; and in the region ${k_1}_T > 39 GeV$ the total EWC
reduce the Born cross section in both approaches, so in the resonance
region (${k_1}_T \sim m_W/2$) the difference between Born and $O(\alpha^3)$
cross sections is $\sim 0.004(0.003)nb/GeV$ for Ref.\cite{pp}(our)
calculations.

The main uncertainty for hadronic part of the EWC in our approach
is the quark mass dependence. From Fig. 6 it can be seen that in the region
${k_1}_T < 40 GeV$ the difference between corrections at
various values of quark masses is rather essential:
so increasing $m_u, m_d$ by a factor of 10 the correction decreases
for $\sim 0.05$. In the vicinity of ${k_1}_T \sim m_W/2$
and for ${k_1}_T > m_W/2$ the discussed difference does not exceed $0.02$.
For the calculation of $O(\alpha^3)$ cross section of Fig.5
and also for the rest analysis we used the values of the current quark
masses of $m_u=5MeV,\ m_d=8MeV$, considering the arguments
concerning the choice of these not well-defined parameters,
which are discussed in Ref.\cite{bom_z} (page 1101, and references therein).
However for quark masses of such an order a quite good correspondence
with the results of calculations by the collinear singularity extraction
method is reached (parameter $\delta_{\theta}\geq 10^{-4}$).

To estimate the scale of radiative effects and it's influence on
observable quantities in the processes (1), the numerical
calculations of the cross sections (Fig.7 and Fig.9), the total
EWC $\delta_{l^{\pm}}^u $  to spin-averaged part of cross section (Fig.8)
\begin{equation}
  \sigma^{\pm}_{EWC}=
    {\bar \sigma_0^{\pm}} (1+\delta_{l^{\pm}}^u )
  + {p_{N}}_2 {\Delta \sigma_0^{\pm}} (1+\delta_{l^{\pm}}^p )
\end{equation}
and the single spin asymmetries
$A^{l^{\pm}}$ taking into consideration EWC (Fig.10)
at typical values for future experiment STAR at the collider RHIC
($\sqrt{S}=500\ GeV$, $-1\leq \eta \leq 2$, $\Delta \Phi=2\pi$)
have been made.
We used the following standard set of electroweak parameters:
$\alpha=1/137.036$,\ $m_W=80.43 GeV$,\ $m_Z=91.19 GeV$,\ $M_H=300.0 GeV$,
the fermion masses:
$ \ m_u=5 MeV$,\ $m_d=8 MeV$,\ $m_s=150 MeV$,\ $m_c=1.5 GeV$,
$\ m_b=4.5 GeV$,\ $m_t=30.0 GeV$;
the GRV94 proton parametrization of parton distribution
functions \cite{GRV94} for unpolarized quarks and
the GRSV96 LO proton parametrization of parton spin densities
\cite{GRSV96}.
We have selected for the $Q^2$ in these $Q^2$-dependent distributions
as well as in Ref.\cite{pp} $Q^2=m_W^2$, since trying to use for $Q$
an explicit momentum transfer in the partonic reaction
(e.g. $Q^2=\hat s$ for Born and final state radiation cross sections)
we have not obtained some noticeable difference for the numerical
estimation of the observable quantities.
And at last, as the contributions to the observables of investigated
processes which have different origin are not distinguishable
experimentally, we did not conduct the comparative analysis for them.

We present in Fig.7 the numerical results for the spin-averaged
part of the double differential cross section as a function of ${k_1}_T$
for the different values of pseudorapidity. As we can see, for $e^+$
and $e^-$ in the final state EWC are essential and increase the Born
cross section in the region ${k_1}_T < m_W/2$ for $\eta =-1,\ 0$; the EWC
are not significant at small ${k_1}_T$ for $\eta >1$,
and in the region ${k_1}_T \geq m_W/2$ for all values $\eta$.
For $\mu^+$ and $\mu^-$ in the final state EWC are most essential
in the resonance region for $\eta=-1,\ 0$, and at the small ${k_1}_T$
in the case $\eta= 1.5,\ 2$ (EWC decrease the Born cross section).

It can be seen from Fig.8, where the corrections $\delta^u$ to
spin-averaged part of Born cross sections are represented,
that muon corrections are lower than that of electron
(difference is approximately $0.12$).
This fact corresponds to a common character of EWC
dependence on masses which regularize the collinear singularity
(the mass $m_{\mu}$ is more than $m_e$ by $2\times 10^2$ times)
and which were discussed in the beginning of this section.
General features of the corrections behavior for all cases
($e^{\pm},\mu^{\pm}$) and for different $\eta$
are the increase in the region ${k_1}_T < m_W/2$, the sharp falling
in the vicinity of resonance and smooth growth at ${k_1}_T > m_W/2$.
The corrections to transverse mass distribution, which are represented
on Fig.9 of Ref.\cite{pp} have a similar behavior.

The scale and the behavior of the radiative corrections to
polarization part of the Born cross section are shown in Fig.9, but
as this part of the cross section is unobservable we shall
not analyze it in details, and consider the influence the radiative
effects to observable quantities -- polarization asymmetries
(\ref{a1}),(\ref{a2}).
So, Fig.10 shows Born asymmetries $A_0$ together with the EWC corrected
results. One can see that the corrections to asymmetries are significant
and decrease the Born value practically in the whole investigated region
except the region of small ${k_1}_T$ and large $\eta$.
Let us remark that the corrections are almost independent of particle mass
in the final state ($m_e$ or $m_{\mu}$).
In other words the ratios of the cross sections (i.e. asymmetries)
are much less sensible to the values of masses than the cross sections
themselves.
For the same reason the corrections to asymmetries do not depend
practically on the choice of quark masses too.
So, if we change the masses of
the light quarks from $m_u=5 MeV,\ m_d=8 MeV$ to the extreme value
$m_u=m_d=0.33 GeV$ we get a shift of radiative corrected asymmetries
by less than $0.01$ in a whole investigated region, and in the region
$\eta=2$, ${k_1}_T=10 GeV$,
which is very important for the quark density analysis at low $x$
(see Section 1)
this shift  equals $0.0032(0.0021)$ for $\mu^-(\mu^+)$ case.
Thus we can suppose that our approach to the calculation
of radiative corrections
to the single spin asymmetries is practically free from
the quark masses uncertainty.

In conclusion we have presented the scheme to use the
processes of the single $W$-boson production
possible at hadron-hadron colliders
for the investigation of the nucleon spin structure.
An approach is proposed
for the determination $u-$ and $d-$ quark (and antiquark) spin densities
at low $x$ by a set of the observable quantities.
The electroweak radiative corrections to the differential cross sections
and the single spin asymmetries of the processes (1) are calculated.
The obtained formulas do not depend on any parameters of the
infrared divergence extraction (e.g. maximum soft photon energy).
We checked that EWC to asymmetries do practically not depend on the masses
of the quarks which regulate in our approach the collinear singularity.
Our results can be used for
studies of nucleon spin at hadron-hadron colliders (RHIC and HERA).
An analysis of the numerical results for the collider experiment STAR
at RHIC shows that the radiative corrections prove to be important.

\section{Acknowledgments}
Author wishes to thank A.N.Ilychev for the assistance in calculation
of the scalar integrals from Appendix A.

\appendix
\section {Phase space of photon and scalar integrals}
\setcounter{equation}{0}

The integral over phase space of the radiated photon can be
presented in the form
\begin{eqnarray}
\label{int}
I[A]&=&\frac{1}{\pi}\int\frac{d^3k}{k_0} \delta[(p_1+p_2-k_1-k)^2][A]
\nonumber \\
&=&\frac{1}{\pi}
     \int \limits _{z^{min}}^{z^{max}}\
     dz
     \int \limits _{{u_1}^{min}}^{{u_1}^{max}}
     \frac{du_1}{\sqrt{R_{u_1}}}[A],
\end{eqnarray}
where
$-R_{u_1}/16=\Delta_4$ is the Gramm determinant,
$ R_{u_1}=-A_{u_1}u_1^2-2B_{u_1}u_1-C_{u_1}$
with the following coefficients:
\begin{eqnarray}
\displaystyle
A_{u_1}=&&(s-v)^2-2(s+v)m^2_l+m_l^4,
\nonumber \\[0.2cm]\displaystyle
B_{u_1}=&&[(s-v)(u-m_2^2)+m_l^2(2v-t-m_1^2+2m_2^2)]v
\nonumber \\[0.2cm]\displaystyle &&
-[(s-m_l^2)(m_2^2-t)+(u+m_1^2-2m_2^2
-2m_l^2)v]z,
\nonumber \\[0.2cm]\displaystyle
C_{u_1}=&&[(v-t+m_2^2)z+(u-m_l^2)v]^2
-m_2^2v[4z^2
\nonumber \\[0.2cm]\displaystyle
&&-2z(s-u-2v+m_1^2-m_l^2)+v(2u-m_2^2+2m_l^2)].
\end{eqnarray}

The limits of the double integration $ u_1^{max/min} $  and $
z^{max/min}$ are the roots of equations
$R_{u_1}=0$ and $u_1^{max}=u_1^{min}$ respectively
$$
u_1^{max/min}=-\frac{B_{u_1}}{A_{u_1}}
   +\!\!/\!\!\!-
\sqrt{(\frac{B_{u_1}}{A_{u_1}})^2-\frac{C_{u_1}}{A_{u_1}}},
$$
$$ z^{max/min}=\frac{1}{2}(\tau +\!\!/\!\!\!- \sqrt{\tau^2-4m_l^2v}),\
   \tau=s-v-m_l^2. $$

During the calculation we used the abbreviations
\begin{eqnarray}
t_{\Gamma w}^2&=&t_w^2+m_W^2\Gamma_W^2,\
t_{\Gamma w}=\sqrt{t_{\Gamma w}^2},
\nonumber\\[0.3cm]
T_u&=&m_W^2(v-u)+su, \
T_{\Gamma u}^{2}=T_u^2+(v-u)^2m_W^2\Gamma_W^2,
\nonumber\\[0.3cm]
T_t&=&m_W^2(v-t)+st, \
T_{\Gamma t}^{2}=T_t^2+(v-t)^2m_W^2\Gamma_W^2.
\end{eqnarray}
and following list of integrals over $u_1$:
\begin{eqnarray}
&&
\displaystyle
J_1=\frac 1{\pi }\int\limits
_{u_1^{min}}^{u_1^{max}}\frac{u_1^2du_1}{\sqrt{R_{u_1}}}
=\frac {3B_{u_1}^2-A_{u_1}C_{u_1}}{2A_{u_1}^{5/2}},\:
\nonumber \\[0.2cm]\displaystyle
&&
\displaystyle
J_2=\frac 1{\pi }\int\limits
_{u_1^{min}}^{u_1^{max}}\frac{u_1du_1}{\sqrt{R_{u_1}}}
=-\frac {B_{u_1}}{A_{u_1}^{3/2}},
\nonumber \\[0.2cm]\displaystyle
&&
\displaystyle
J_3=\frac 1{\pi }\int\limits _{u_1^{min}}^{u_1^{max}}\frac{du_1}{\sqrt{R_{u_1}}}
=\frac 1{\sqrt{A_{u_1}}},\;
\nonumber \\[0.2cm]\displaystyle
&&
\displaystyle
J_4=\frac 1{\pi }\int\limits
_{u_1^{min}}^{u_1^{max}}\frac{du_1}{u_1\sqrt{R_{u_1}}}
=\frac 1{\sqrt{C_{u_1}}},
\nonumber \\[0.2cm]\displaystyle
&&
\displaystyle
J_5=\frac 1{\pi }\int\limits
_{u_1^{min}}^{u_1^{max}}\frac{du_1}{u_1^2\sqrt{R_{u_1}}}
=-\frac {B_{u_1}}{C_{u_1}^{3/2}},
\nonumber \\[0.2cm]\displaystyle
&&
\displaystyle
J_6=\frac 1{\pi }\int\limits
_{u_1^{min}}^{u_1^{max}}\frac{du_1}{z_1\sqrt{R_{u_1}}}
=\frac 1{\sqrt{C_{z_1}}},
\nonumber \\[0.2cm]\displaystyle
&&
\displaystyle
J_7=\frac 1{\pi }\int\limits
_{u_1^{min}}^{u_1^{max}}\frac{du_1}{z_1^2\sqrt{R_{u_1}}}
=-\frac {B_{z_1}}{C_{z_1}^{3/2}},
\end{eqnarray}
where
\begin{eqnarray}
B_{z_1}=-A_{u_1}(v+z)-B_{u_1},\;
\nonumber \\[0.6cm]\displaystyle
C_{z_1}=A_{u_1}(v+z)^2+2B_{u_1}(v+z)+C_{u_1}.
\end{eqnarray}

At last, we list the following table of scalar integrals ($W$-boson
width is neglected where it is possible)
\begin{eqnarray}
&&I_1=I[1]=1,
\nonumber \\[0.3cm] \displaystyle
&&I_2=I[\frac1 z]=\frac {\log [(s-v)^2/m_l^2v]}{s-v},
\nonumber \\[0.3cm] \displaystyle
&&I_3=I[\frac 1 {u_1}]=\frac {\log [(s+u)^2/m_2^2v]}{s+u},\;
\nonumber \\[0.3cm] \displaystyle
&&I_4=m_l^2I[\frac{u_1}{z^2}]=-\frac {u}{s-v}, \;
\nonumber \\[0.3cm]\displaystyle
&&I_5=m_2^2I[\frac 1 {u_1^2}]=\frac 1v,\;
\nonumber \\[0.3cm] \displaystyle
&&I_6=I[u_1]=\frac 12(s+u), \;
\nonumber \\[0.3cm]\displaystyle
&&I_7=I[\frac {u_1} z]=\frac {uv(1-\log
[(s-v)^2/m_l^2v])-st}{(s-v)^2},
\nonumber \\[0.3cm]\displaystyle
&&I_8=I[\frac 1 {zz_1}]=-\frac 1{vt}\log
\frac{t^2}{m_1^2m_l^2},\;\;
\nonumber \\[0.3cm] \displaystyle
&&I_{9}=I[\frac 1 {zu_1}]=-\frac 1{vu}\log \frac{u^2}{m_2^2m_l^2},
\nonumber \\[0.3cm]\displaystyle
&&I_{10}=I[\frac {u_1^2}z]=
\frac {u^2v^2}{(s-v)^3}\log [\frac {(s-v)^2}{m_l^2v}]
\nonumber \\[0.3cm]\displaystyle
&&\qquad \qquad
+\frac {s^2t^2+4uvst-3u^2v^2}{2(s-v)^3},
\nonumber \\[0.3cm]\displaystyle
&&I_{11}=m_l^2I[\frac{u_1^2}{z^2}]=\frac {vu^2}{(s-v)^2},\;
\end{eqnarray}
\ \\
\begin{eqnarray}
&&I_{12}=Re I[\Pi_q ]=-\frac 1{(s-v)} \log \Bigl[\frac
{t_{\Gamma w}}{m_W^2}\Bigr],
\nonumber \\[0.3cm]\displaystyle
&&I_{13}=Re I[\frac {\Pi_q}{z_1}]=\frac {T_u}{T_{\Gamma u}^2}
\log \Bigl[\frac {T_{\Gamma
u}^2}{m_1^2m_W^2t_{\Gamma w}v}\Bigr],
\nonumber \\[0.3cm]\displaystyle
&&I_{14}=Re I[u_1\Pi_q]=
\frac {(uv-st)}{(s-v)^2}
+ \frac {(m_w^2uv-stt_w)}{(s-v)^3}\log
\Bigl[\frac{t_{\Gamma w}}{m_W^2}\Bigr],
\nonumber \\[0.3cm]\displaystyle
&&I_{15}=Re I[u_1^2\Pi_q]=\frac{3s^2t^2-4stuv-u^2v^2}{2(s-v)^3}
\nonumber \\[0.3cm]\displaystyle
&&\qquad
-\frac{s^2t^2-4stuv+u^2v^2}{(s-v)^4}m_W^2
\nonumber \\[0.3cm]\displaystyle
&&\qquad
+\frac {2m_W^2t_wstvu-(stt_w-m_W^2uv)^2}{(s-v)^5}
\log \Bigl[ \frac
{t_{\Gamma w}}{m_W^2}\Bigr],
\nonumber \\[0.3cm]\displaystyle
&&I_{16}=Re I[\frac {\Pi_q }{u_1}]=I_{13}\{ m_1\rightarrow m_2,
u\leftrightarrow t\},
\nonumber \\[0.3cm]\displaystyle
&&I_{17}=m_2^2 Re I[\frac {\Pi _q}{u_1^2}]=\frac
{(v-t)T_t}{vT_{\Gamma t}^2},\;
\end{eqnarray}
\ \\
\begin{eqnarray}
&&I_{18}=I[\Pi _q\Pi _q^+]=
\frac{ \pi /2-\arctan (t_w/(m_W\Gamma _W) )-\arctan (\Gamma _W/m_W)}
{m_W\Gamma _W (s-v)},
\nonumber \\[0.3cm]\displaystyle
&&I_{19}=I[\frac 1{z_1}\Pi _q\Pi _q^+]=
\frac{1}{T_{\Gamma u}^2}\Biggl [ -\frac {T_u}{m_W\Gamma
_W}\Bigl (
\arctan\Bigl [\frac{(s-m_W^2)m_W}{s\Gamma_W}\Bigr ]
\nonumber \\[0.3cm]\displaystyle
&&-\arctan\Bigl [\frac{(s-m_W^2)T_u}{vsm_W\Gamma _W} \Bigr ] \Bigr )
  +(s+t)\log \Bigl[ \frac{|u|\sqrt{T^2_{\Gamma_u}}}{m_1^2m_W^2v}\Bigr]
\Biggr ],
\nonumber \\[0.3cm]\displaystyle
&&I_{20}=I[\frac 1 {u_1}\Pi _q\Pi _q^+]=I_{19}\{m_1^2 \rightarrow
m_2^2,t\leftrightarrow u\},
\nonumber \\[0.3cm]\displaystyle
&&I_{21}=m_1^2I[\frac 1{z_1^2}(\Pi _q\Pi _q^+-\Pi _l\Pi _l^+)]=
(s(v-2u)+2m_W^2(u-2v))s\Pi _l\Pi _l^+/T_{\Gamma u}^2,
\nonumber \\[0.3cm]\displaystyle
&&I_{22}=m_2^2I[{\frac 1{u_1^2}}\Pi _q\Pi _q^+]=
\frac {(s+u)^2}{vT_{\Gamma t}^2},
\nonumber \\[0.3cm]\displaystyle
&&I_{23}=I[u_1\Pi _q\Pi _q^+]=
\frac 1 {(s-v)^3}  \Biggl [
(st-uv)\log \frac {t_{\Gamma w}}{m_W^2}
\nonumber \\[0.3cm]\displaystyle
&&
+\frac{t_wst-m_W^2uv}{ m_W^2}\Biggl(\frac{m_W}{\Gamma _W}\Bigl(\pi/2
-\arctan \Bigl [\frac{t_w}{\Gamma _W m_W}\Bigr]\Bigr)-1
\Biggr ) \Biggr ] ,
\nonumber \\[0.3cm]\displaystyle
&&I_{24}=I[u_1^2\Pi _q\Pi _q^+]=\frac 1 {(s-v)^5}
\Biggl[\frac {
(\pi-2\arctan [t_w/(m_W\Gamma _W)])
}{4 m_W\Gamma _W}
\nonumber \\[0.3cm]\displaystyle &&
\times [3(stt_w-m_W^2uv)^2
-(st+m_W^2(s+u))^2(s-v)^2]
\nonumber \\[0.3cm]\displaystyle &&
+s^2t^2\Bigl(s-v-\frac{t_w^2}{m_W^2}\Bigr)
-4(s-v-t_w)stuv-t_wu^2v^2
\nonumber \\[0.3cm]\displaystyle &&
+[t_ws^2t^2-2stuv(t_w+m_W^2)+m_W^2u^2v^2]
\log\Bigl[\frac{t_{\Gamma w}^2}{m_W^4}
\Bigr] \Biggr],
\nonumber \\[0.3cm]\displaystyle
&&I_{25}=I[\frac{\Pi _q\Pi _q^+}{z_1(z+v)}]
=\frac 1{T_{\Gamma u}^2}
\Biggl [
\frac{(v-u)^2}{sv}\log \frac s{m_1^2}
\nonumber \\[0.3cm]\displaystyle &&
+  (vs-2(s+t)(s-m_W^2))\Pi _l\Pi _l^+
   \log \frac {vm_W^2}{ \sqrt{T^2_{\Gamma_u}}}
\nonumber \\[0.3cm]\displaystyle &&
  -\frac{vs-(s+t)(s-m_W^2)}{m_W\Gamma_W}(s-m_W^2)\Pi _l\Pi _l^+
\nonumber \\[0.3cm]\displaystyle &&
 \times  \Bigl ( \arctan [ \frac{(s-m_W^2)m_W}{s\Gamma _W}    ]
         - \arctan [ \frac{(s-m_W^2)T_u}{vsm_W\Gamma _W} ]
\Bigl )
\Biggl ],
\nonumber \\[0.3cm]\displaystyle
&&I_{26}=I[\frac{\Pi _q\Pi _q^+}{u_1(z+v)}]=I_{25}\{m_1^2\rightarrow
m_2^2, t \leftrightarrow u\},
\nonumber \\[0.3cm]\displaystyle
&&I_{27}=I[\frac 1{z_1u_1}]=\frac 1{sv}\log \frac {s^2}{m_1^2m_2^2},
\nonumber \\[0.3cm]\displaystyle
&&I_{28}=I[\frac 1{z_1}]=\frac 1{v-u}\log \frac
{(v-u)^2}{m_1^2v}.
\end{eqnarray}

\section{ Expressions for the $V_j$.}
\setcounter{equation}{0}
\subsection{The case of  $q \bar{q} \rightarrow l^- \bar{\nu}_l,
                          \bar{q} q \rightarrow l^+ \nu_l$
            subprocesses}
\begin{equation}
V_l = V_l^* = 4uI_4-I_6-uI_7+I_{10}-2I_{11}
\end{equation}
\begin{equation}
\begin{array}{lll}
\displaystyle
V_{lq}=& V_{lq}^* =
Q_i\{-(2u(u-t)+vt)I_2+(u - t)(I_7-I_{14})
\\[0.2cm] \displaystyle &
+ vt( 2\Pi_lu^2-2u + v)I_8 -(t_w(s+u) -vt-2u(u-t))I_{12}
\\[0.2cm] \displaystyle &
+(t_w(v-u)(v-2u)-t(2u^2-2uv+v^2))I_{13}\}/t_{w\Gamma}
\\[0.2cm] \displaystyle &
+Q_{i'}\{2t_wI_1 + 4u^2I_2
- 2t_wuI_3- 2uI_7- 2\Pi_lu^3vI_9
\\[0.2cm] \displaystyle &
-2(t_w-u)((t_w - 2u)I_{12}+ I_{14}-u(t_w-u)I_{16})\}/t_{w\Gamma}
\end{array}
\end{equation}
\begin{equation}
\begin{array}{lll}
\displaystyle
V_q=&
-Q_i^2u\{vI_{19}+2uI_{21}\}+Q_{i'}^2\{-I_3+2(\Pi_l\Pi_l^+u^2-1)I_5+I_{12}
\\[0.2cm] \displaystyle &
+(2t_w-u)I_{16}+(t_w - u)(4I_{17}-t_wI_{20}-2(t_w-u)I_{22})
\\[0.2cm] \displaystyle &
-t_wI_{18} \} +Q_iQ_{i'}\{(s - u)I_{16} -(s - v)I_{18}
\\[0.2cm] \displaystyle &
- (2u^2- 3uv + v^2)I_{19}-((s-u)(t_w-2u)+sv)I_{20}
\\[0.2cm] \displaystyle &
+sv(v-2u)(I_{25}+I_{26})+ 2su^2(I_{25}+I_{26}-\Pi _l\Pi_l^+I_{27})\}
\end{array}
\end{equation}
\begin{equation}
\begin{array}{lll}
\displaystyle
V_{lw}=&
\{st_wI_1+2su^2I_2+2t_wI_6-u(2s-v)I_7+(s-v)I_{10}
\\[0.2cm] \displaystyle &
-(s(t_w-u)^2+u(t_w(u-v)+su))I_{12}
\\[0.2cm] \displaystyle &
-(t_w-u)(2t_w+2s -v)I_{14}- (s + 2t_w - v)I_{15}\}/t_{w\Gamma}
\end{array}
\end{equation}
\begin{equation}
\begin{array}{lll}
\displaystyle
V_{qw}=&
Q_i\{sI_{12}-(2u^2-3uv+v^2)I_{13}+(s(2u-t_w)+u^2+vt))I_{18}
\\[0.2cm] \displaystyle &
+ ((t_w-v+s)(2u^2 - 3uv  + v^2)+suv)I_{19}+tI_{23}\}
\\[0.2cm] \displaystyle &
+Q_{i'}\{-2I_1 -(s-u)I_3+(2s+4t_w-4u-v)I_{12}+2I_{14}
\\[0.2cm] \displaystyle &
+(uv+2(t_w-u)(s-u))I_{16}
\\[0.2cm] \displaystyle &
+(t_w(4u+v-2t_w-2s)+u(2s-u-3v))I_{18}
\\[0.2cm] \displaystyle &
-(t_wuv +(s-u)t_w(t_w-2u)+2u^2(s-v))I_{20}
\\[0.2cm] \displaystyle &
+(t-2t_w+2u)I_{23}\}
\end{array}
\end{equation}
\begin{equation}
\begin{array}{lll}
\displaystyle
V_w=&
-sI_1-I_6+(2st_w-u(3s+t))I_{12}+(2s+2t_w-u-v)I_{14}
\\[0.2cm] \displaystyle &
+I_{15}+( t_w(u(3s+t)-st_w)
+ u(uv  -v^2-2su))I_{18}
\\[0.2cm] \displaystyle &
+(u(2s-v)+t_w(u +v-t_w-2s))I_{23}- (s-v+t_w)I_{24}
\end{array}
\end{equation}

\subsection{The case of for $q \bar{q} \rightarrow l^+ \nu_l,
                             \bar{q} q \rightarrow l^- \bar{\nu}_l$
            subprocesses}
\begin{equation}
V_l=V_l^*(t\leftrightarrow u)
\end{equation}
\begin{equation}
V_{lq}=V_{lq}^*(
t\leftrightarrow u,\; m_1\leftrightarrow m_2,\; Q_i\leftrightarrow Q_{i'})
\end{equation}
\begin{equation}
\begin{array}{lll}
\displaystyle
V_q=&
Q_i^2\{I_{12}+(2t_w-t)I_{13}-t_wI_{18}+t_w(t-t_w)I_{19}
\\[0.2cm] \displaystyle &
-2(t-t_w)^2I_{21}
-I_{28}\} +Q_{i'}^2t\{2t\Pi_l\Pi_l^+I_5-vI_{20}
-2tI_{22}\}
\\[0.2cm] \displaystyle &
+Q_iQ_{i'}\{(s-t)I_{13}
-(s-v)I_{18}
\\[0.2cm] \displaystyle &
+((2t-t_w)(s-t)-sv)I_{19}
-(2t^2-3tv +v^2)I_{20}
\\[0.2cm] \displaystyle &
+(2t^2-2tv+v^2)s(I_{25}+I_{26}-\Pi_l\Pi_l^+I_{27})
\\[0.2cm] \displaystyle &
+\Pi_l\Pi_l^+sv(v-2t)I_{27}\}
\end{array}
\end{equation}
\begin{equation}
\begin{array}{lll}
\displaystyle
V_{lw}=&
\{2t_w(t-v)I_1+(2st(t-v)-v^2u)I_2+2t_wI_6
\\[0.2cm] \displaystyle &
+(2st+tv+2vu)I_7+(s-v)I_{10}
\\[0.2cm] \displaystyle &
+((u-3t_w)v^2+2(st+t_w^2)(v-t)+tt_w(4v-t))I_{12}
\\[0.2cm] \displaystyle &
+(2v(3t_w-u)-(t+t_w)(2t_w+v)-2st)I_{14}
\\[0.2cm] \displaystyle &
-(s-v+2t_w)I_{15}\}/t_{w\Gamma}
\end{array}
\end{equation}
\begin{equation}
\begin{array}{lll}
\displaystyle
V_{qw}=&
Q_i\{(s-3t+2v)I_{12}+(2(s-t)(t_w-t)+tv)I_{13}-2I_{14}
\\[0.2cm] \displaystyle &
+(t(2s-t-v)-t_w(s-3t+2v)+vu)I_{18}
\\[0.2cm] \displaystyle &
+(t(2s+t_w-v)(t_w-t)-st_w^2+t^2(v-t_w))I_{19}
\\[0.2cm] \displaystyle &
+(2t_w-2t-u)I_{23}-(s-t)I_{28}\} + Q_{i'}\{(v-t)I_{12}
\\[0.2cm] \displaystyle &
-(2t^2-3tv+v^2)I_{16} + (t(2s+t+t_w)+v(2u-t_w))I_{18}
\\[0.2cm] \displaystyle &
+((s-v+t_w)(2t^2+v^2)-tv(2s+3(t_w-v)))I_{20}-uI_{23}\}
\end{array}
\end{equation}
\begin{equation}
\begin{array}{lll}
\displaystyle
V_w=&
(v-t)I_1-I_6+(2t_w(t-v)+t^2-3tv+2v^2)I_{12}
\\[0.2cm] \displaystyle &
+(t+2t_w-3v)I_{14}+I_{15}
\\[0.2cm] \displaystyle &
+((v-t-t_w)(v-t)(v-t_w)-s(2t(t-v)+v^2))I_{18}
\\[0.2cm] \displaystyle &
+((2s+t)(v-t)-t_w^2+3t_wv -2v^2)I_{23} -(s-v+t_w)I_{24}
\end{array}
\end{equation}

\newpage
\input{epsf}

\begin{figure}
\vspace{20mm}
\begin{tabular}{cc}
\begin{picture}(60,60)
\put(-20,-40){
\epsfxsize=7cm
\epsfysize=7cm
\epsfbox{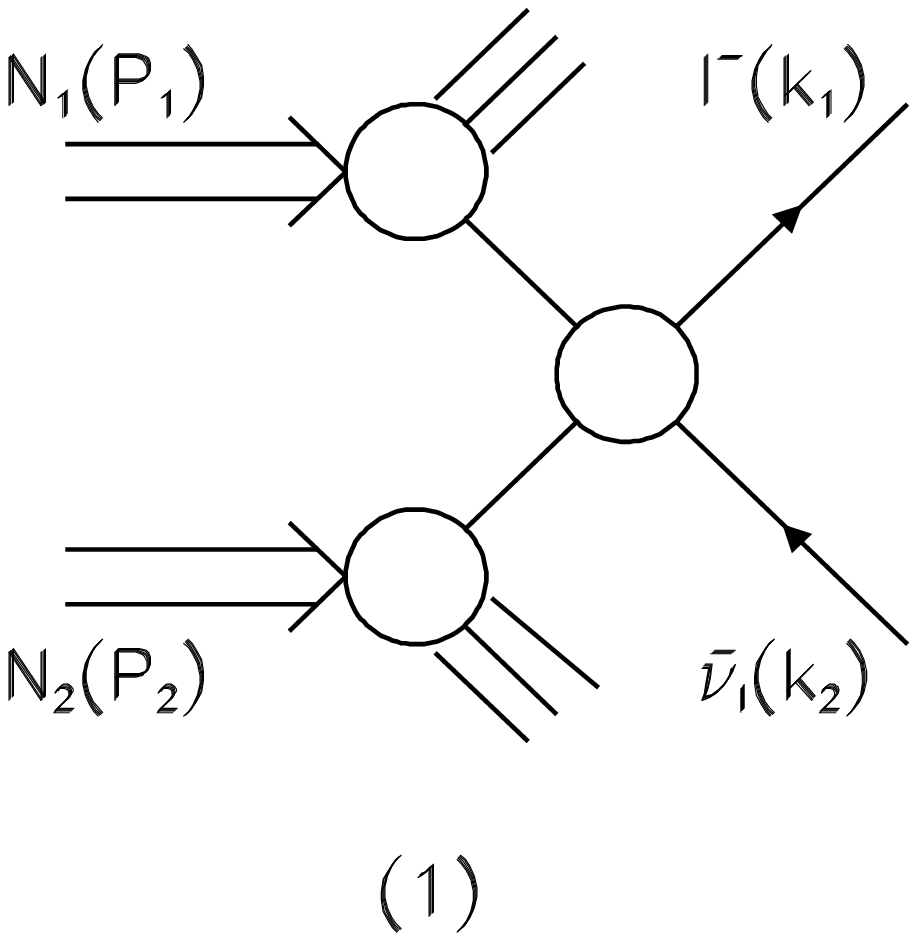} }
\end{picture}
&
\begin{picture}(60,100)
\put(80,-40){
\epsfxsize=7cm
\epsfysize=7cm
\epsfbox{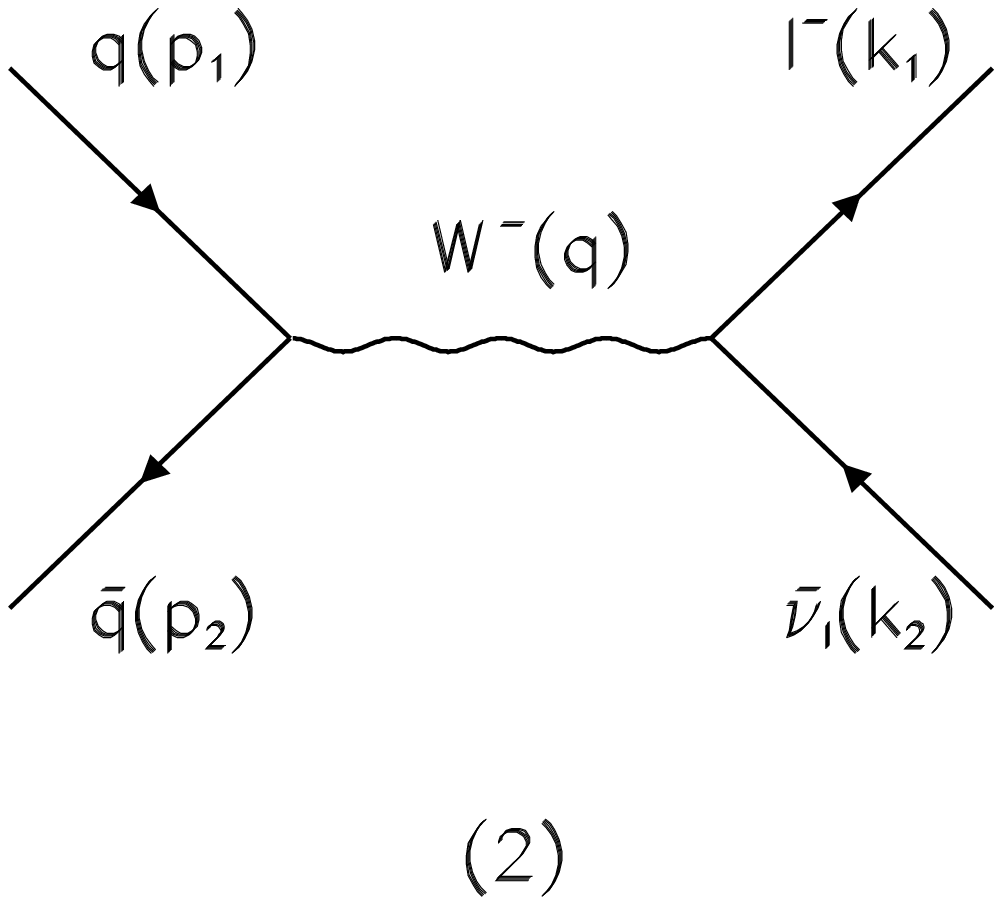} }
\end{picture}
\end{tabular}
\caption{\protect\it
(1) Sketch illustrating the interaction of parton
1 and 2 from incident nucleons $N_1$ and $N_2$, respectively.
The partons carry fractional momenta $x_1$ and $x_2$ and
interact to produce a charged lepton and antineutrino.
(2) The diagram for the lowest order subprocess
$q\bar q\rightarrow l^- \bar \nu$.}
\label{1f}
\end{figure}

\vspace{85mm}

\begin{figure}
\vspace{50mm}
\begin{tabular}{c}
\begin{picture}(60,60)
\put(0,-60){
\epsfxsize=7cm
\epsfysize=7cm
\epsfbox{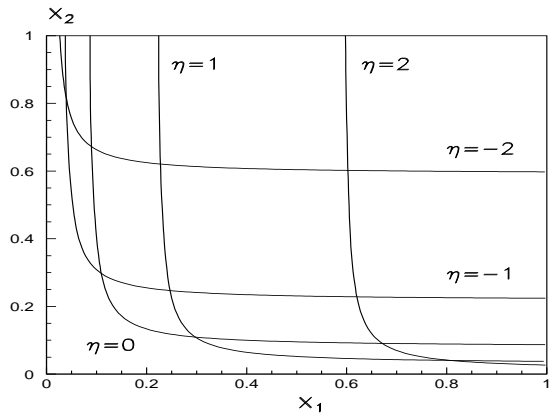} }
\end{picture}
\end{tabular}
\vspace*{25mm}
\caption{\protect\it
Physically allowed region of variables $x_1$ and $x_2$. Curves on the
plot are $x_2^0$ as a function of $x_1$ at different $\eta$
(kinematics of RHIC
experiment STAR: $\sqrt{S}$=500 GeV, ${k_1}_{\perp}$=40 GeV).
}
\label{2}
\end{figure}


\begin{figure}
\vspace{10mm}
\begin{tabular}{ccc}
\begin{picture}(60,60)
\put(-20,0){
\epsfxsize=5cm
\epsfysize=5cm
\epsfbox{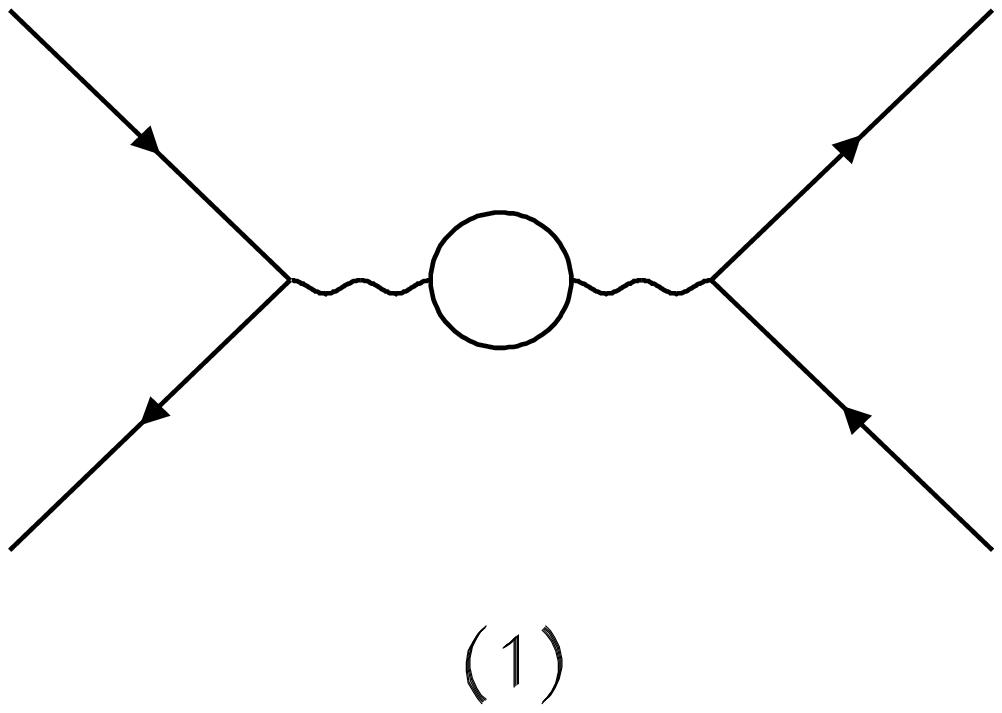} }
\end{picture}
&
\begin{picture}(60,100)
\put(20,0){
\epsfxsize=5cm
\epsfysize=5cm
\epsfbox{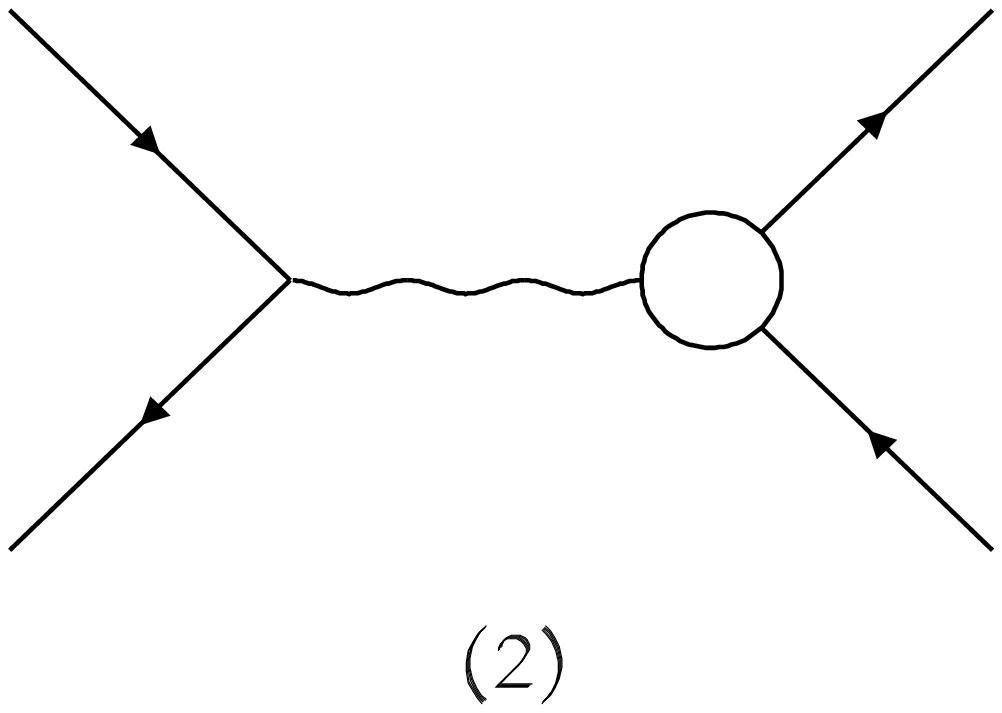} }
\end{picture}
&
\begin{picture}(60,100)
\put(60,0){
\epsfxsize=5cm
\epsfysize=5cm
\epsfbox{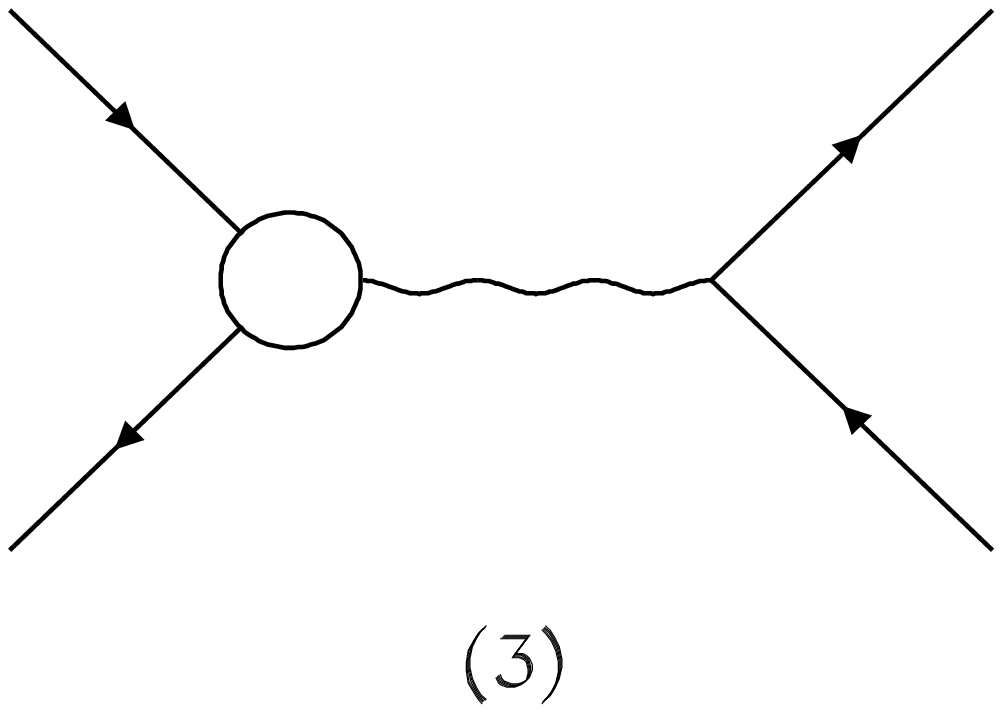} }
\end{picture}
\\[-5mm]
\begin{picture}(60,60)
\put(-20,0){
\epsfxsize=5cm
\epsfysize=5cm
\epsfbox{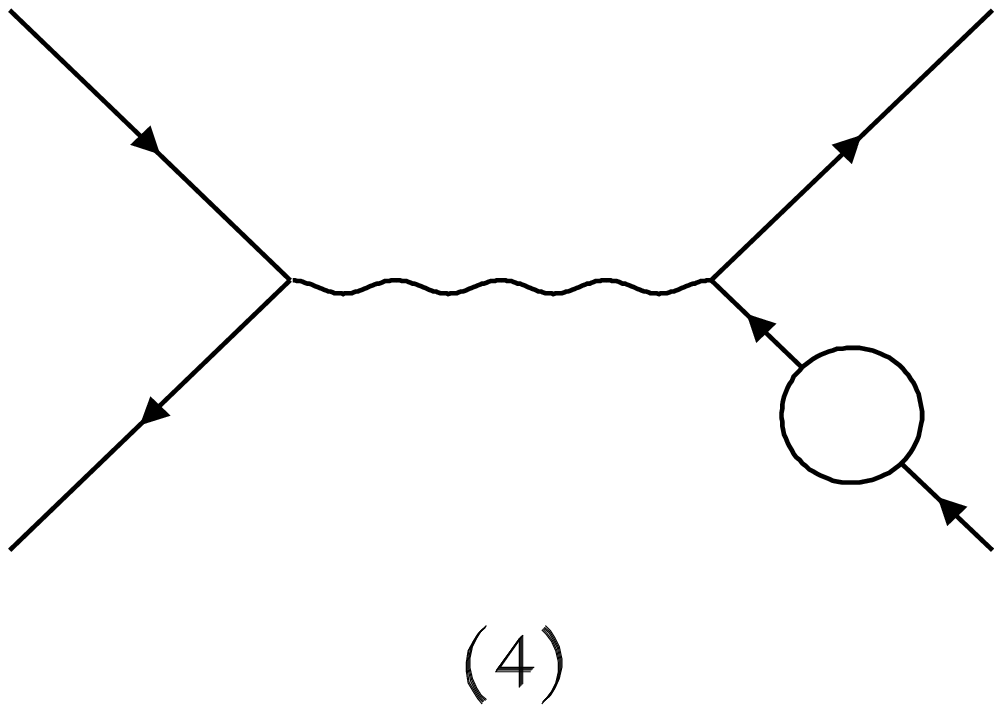} }
\end{picture}
&
\begin{picture}(60,100)
\put(20,0){
\epsfxsize=5cm
\epsfysize=5cm
\epsfbox{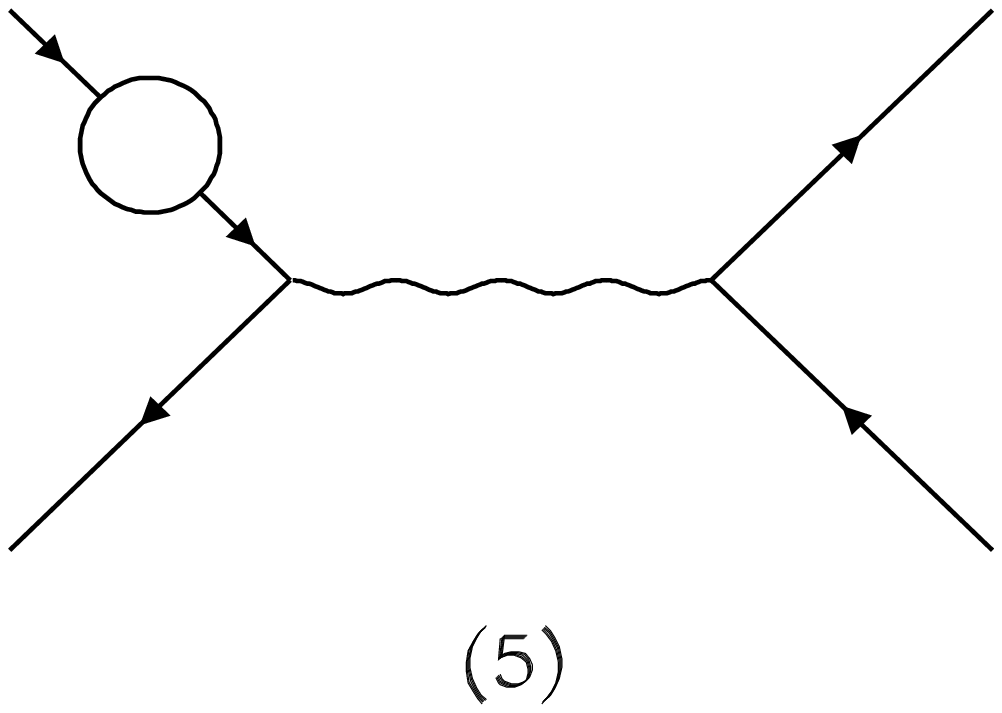} }
\end{picture}
&
\begin{picture}(60,100)
\put(60,0){
\epsfxsize=5cm
\epsfysize=5cm
\epsfbox{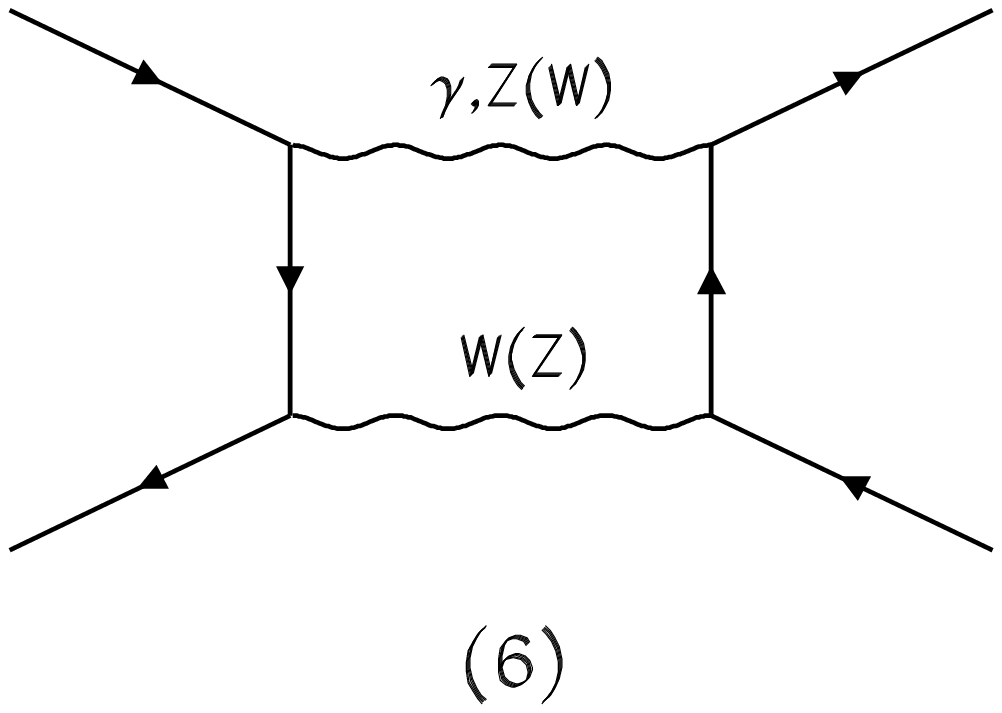} }
\end{picture}
\\[7mm]
\begin{picture}(60,60)
\put(-20,0){
\epsfxsize=5cm
\epsfysize=5cm
\epsfbox{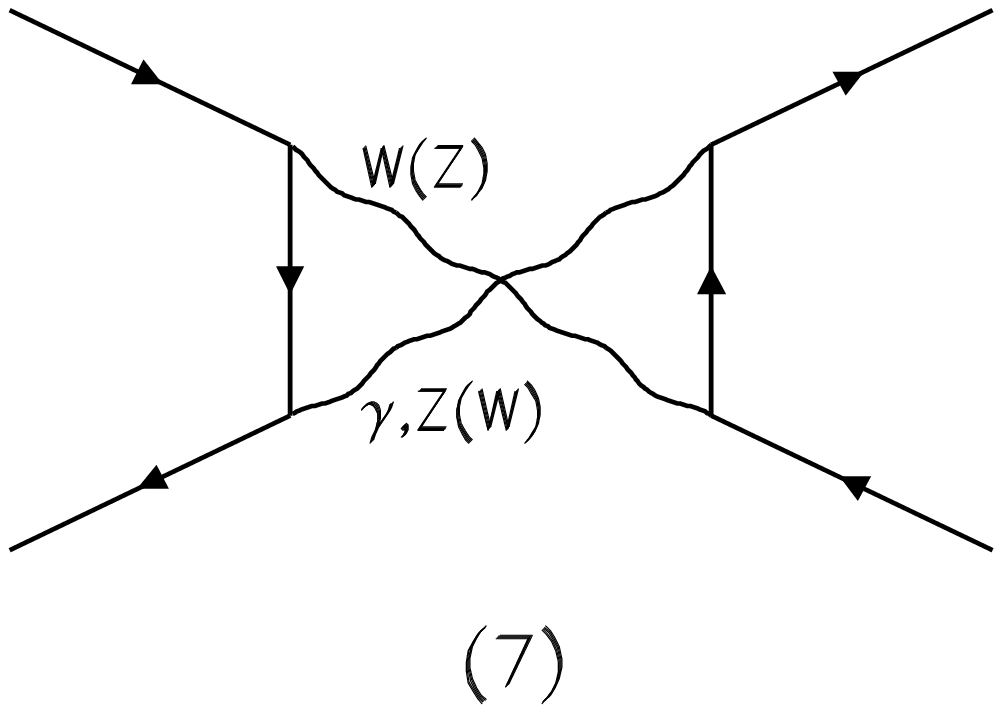} }
\end{picture}
\end{tabular}
\vspace*{-15mm}
\caption{\protect\it
The virtual one-loop diagrams for $q \bar q \rightarrow l^- \bar \nu$
process. The contributions to the self-energies and vertex corrections
are symbolized by the empty loops, an explicit representation
can be found in Ref.\cite{BS2}.}
\label{3}
\end{figure}


\begin{figure}
\vspace{25mm}
\begin{tabular}{ccc}
\begin{picture}(60,60)
\put(-20,0){
\epsfxsize=5cm
\epsfysize=5cm
\epsfbox{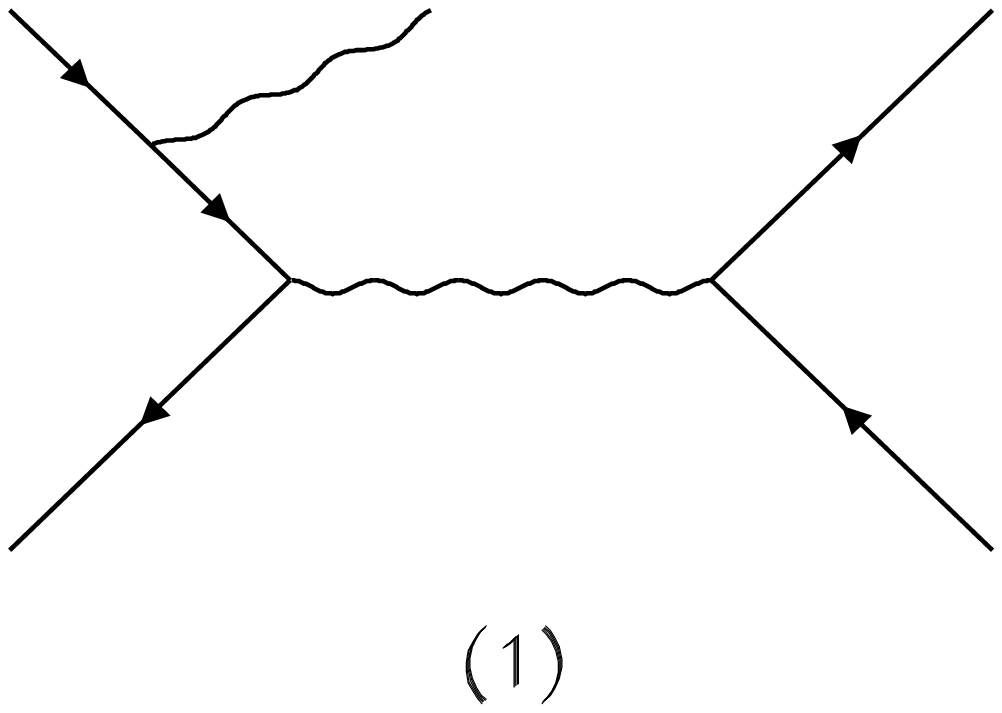} }
\end{picture}
&
\begin{picture}(60,100)
\put(20,0){
\epsfxsize=5cm
\epsfysize=5cm
\epsfbox{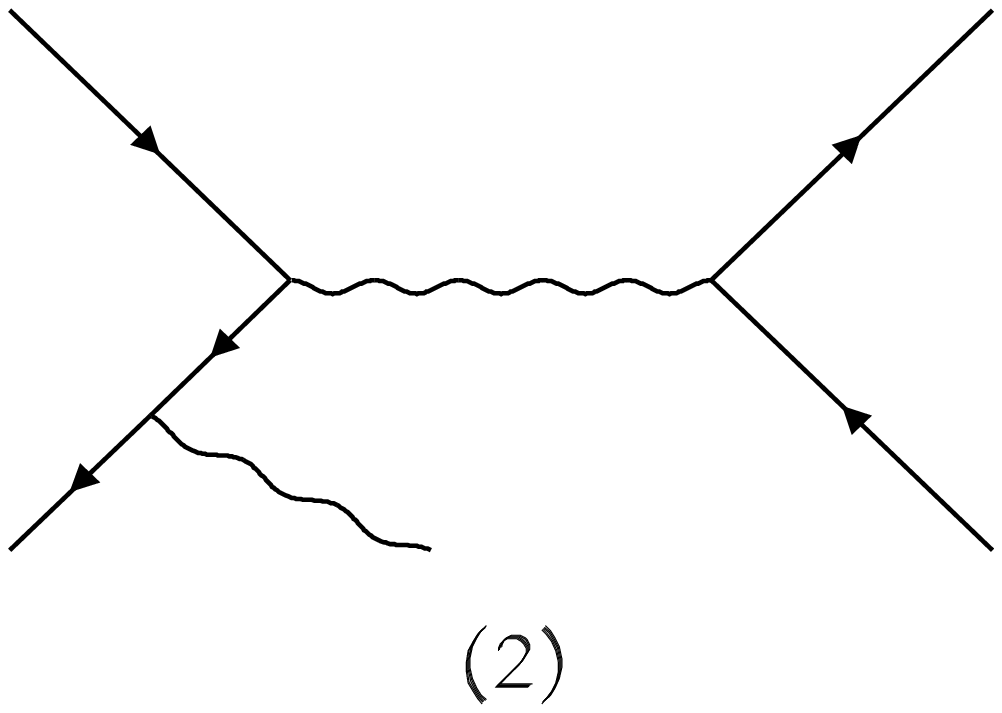} }
\end{picture}
&
\begin{picture}(60,100)
\put(60,0){
\epsfxsize=5cm
\epsfysize=5cm
\epsfbox{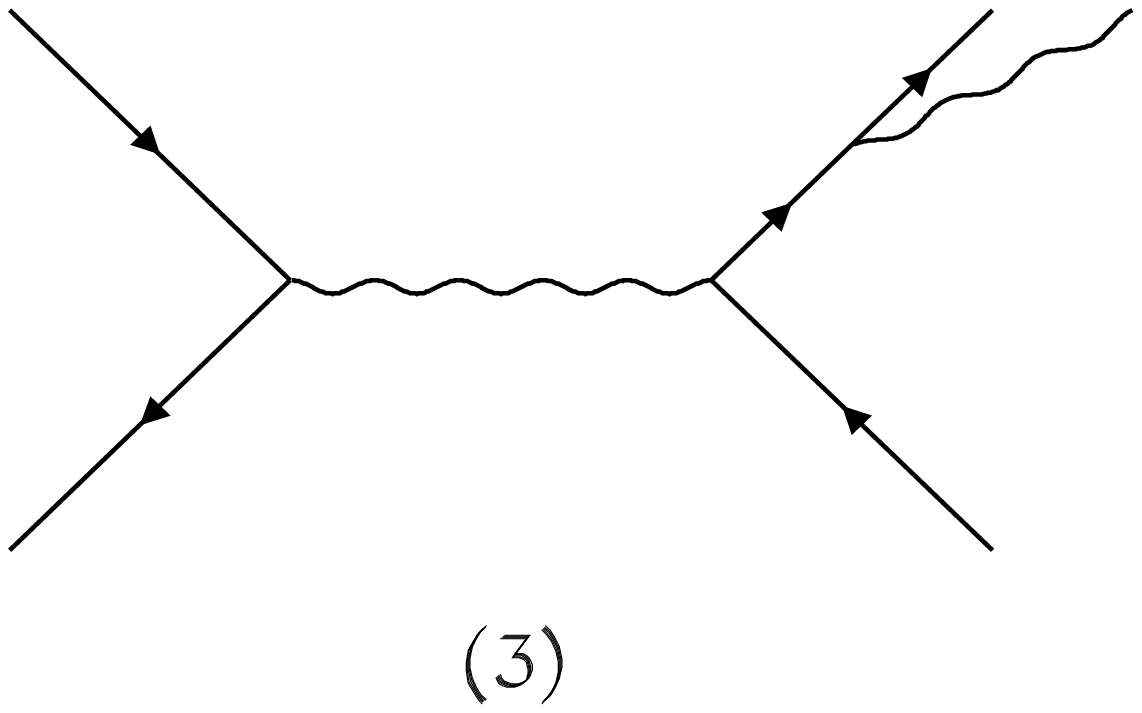} }
\end{picture}
\\[5mm]
\begin{picture}(60,60)
\put(-20,0){
\epsfxsize=5cm
\epsfysize=5cm
\epsfbox{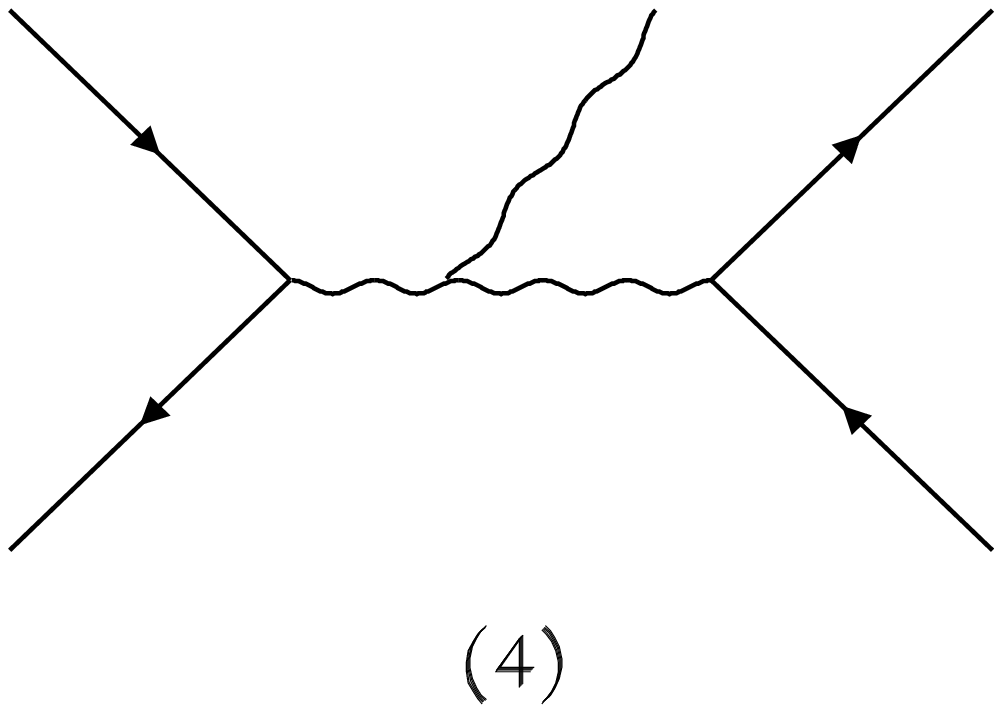} }
\end{picture}
\end{tabular}
\vspace*{-15mm}
\caption{\protect\it
Bremsstrahlung diagrams for $q \bar q \rightarrow l^- \bar \nu \gamma$
process.
}
\label{4}
\end{figure}


\begin{figure}
\vspace{20mm}
\begin{tabular}{c}
\begin{picture}(60,60)
\put(0,-205){
\epsfxsize=9cm
\epsfysize=9cm
\epsfbox{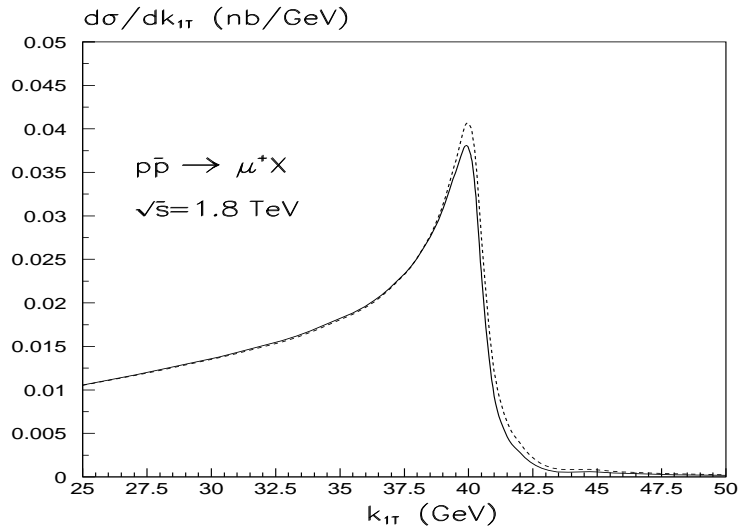} }
\end{picture}
\end{tabular}
\vspace*{55mm}
\caption{\protect\it
Differential cross sections for $p\bar p \rightarrow \mu^+X$
at $\sqrt{S}=1.8$ TeV, $-1.2\leq \eta \leq 1.2$, $\Delta \Phi=2\pi$
(Tevatron) as a function of ${k_1}_T$. Shown is the
muon transverse momentum spectrum in the Born approximation
(dashed line) and taking into account the total EWC (solid line).
We use the MRS LO 98 set of parton distribution functions \cite{MRS98}.
}
\label{5}
\end{figure}


\begin{figure}
\vspace{15mm}
\begin{tabular}{c}
\begin{picture}(60,60)
\put(0,-225){
\epsfxsize=9cm
\epsfysize=9cm
\epsfbox{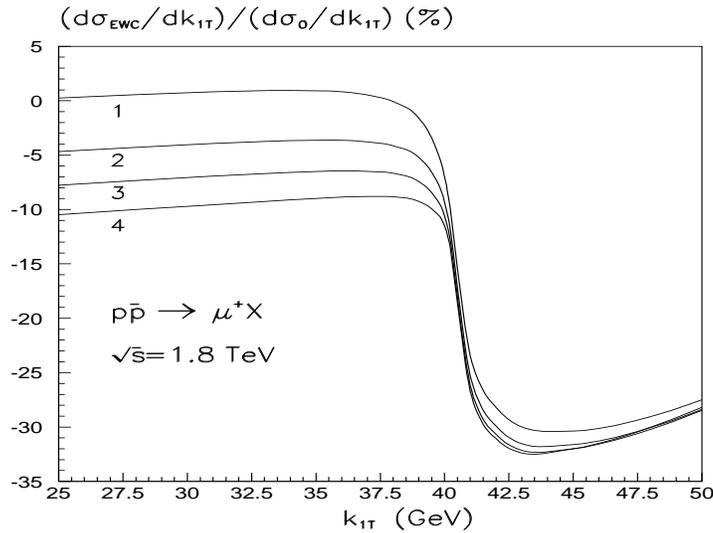} }
\end{picture}
\end{tabular}
\vspace*{60mm}
\caption{\protect\it
The ratio of the $O(\alpha^3)$ and the Born cross sections
as a function of ${k_1}_T$ at different values of
quark masses:
$m_u$=5 $MeV$, $m_d$=8 $MeV$ (curve 1),
$m_u$=30 $MeV$, $m_d$=30 $MeV$ (curve 2),
$m_u$=100 $MeV$, $m_d$=100 $MeV$ (curve 3),
$m_u$=0.33 $GeV$, $m_d$=0.33 $GeV$ (curve 4).
The rest of the parameters is identical to that in Fig.5.
}
\label{6}
\end{figure}


\begin{figure}
\vspace{25mm}
\begin{tabular}{cc}
\begin{picture}(60,60)
\put(-20,-40){
\epsfxsize=6cm
\epsfysize=6cm
\epsfbox{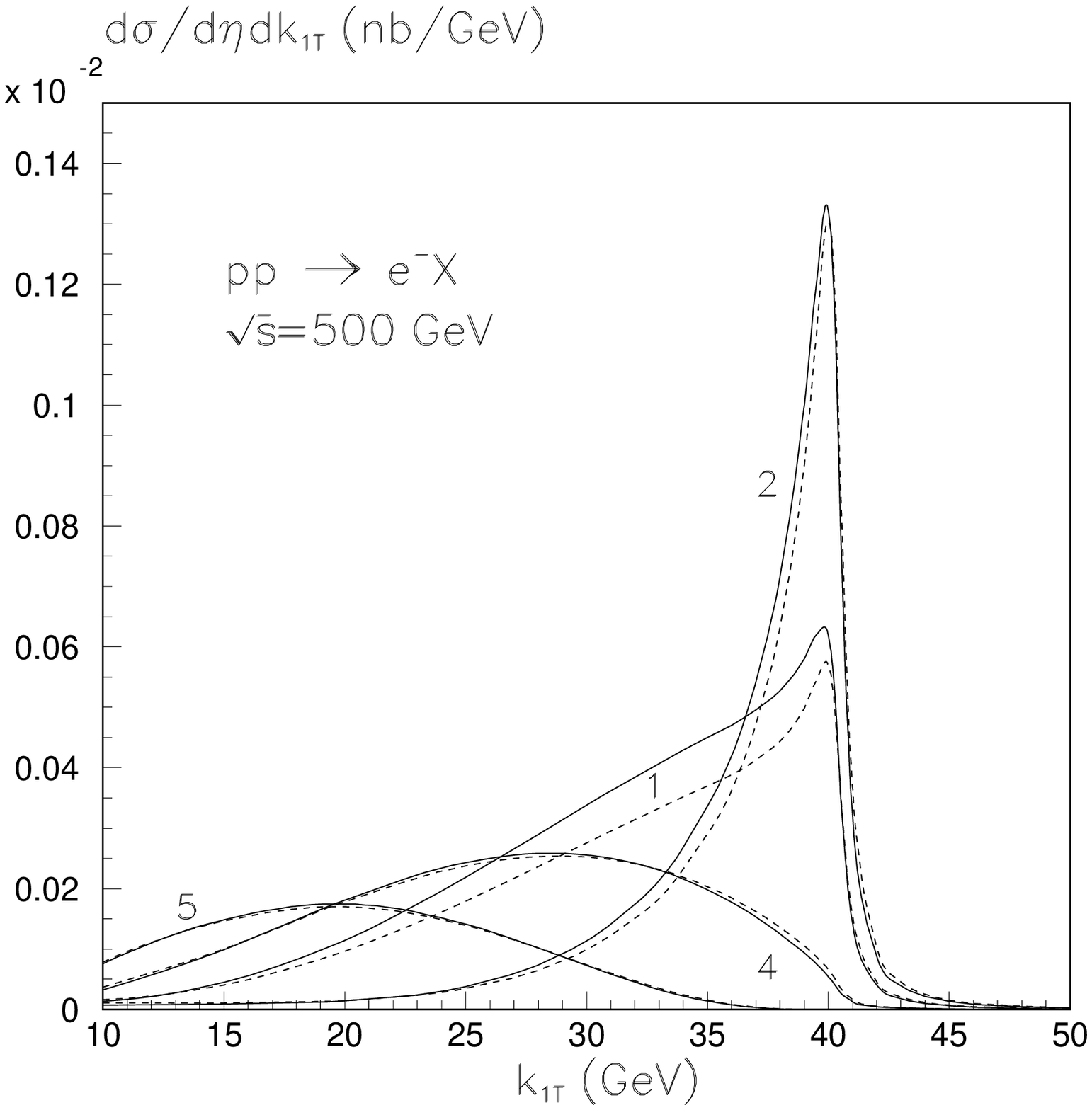} }
\end{picture}
&
\begin{picture}(60,100)
\put(110,-40){
\epsfxsize=6cm
\epsfysize=6cm
\epsfbox{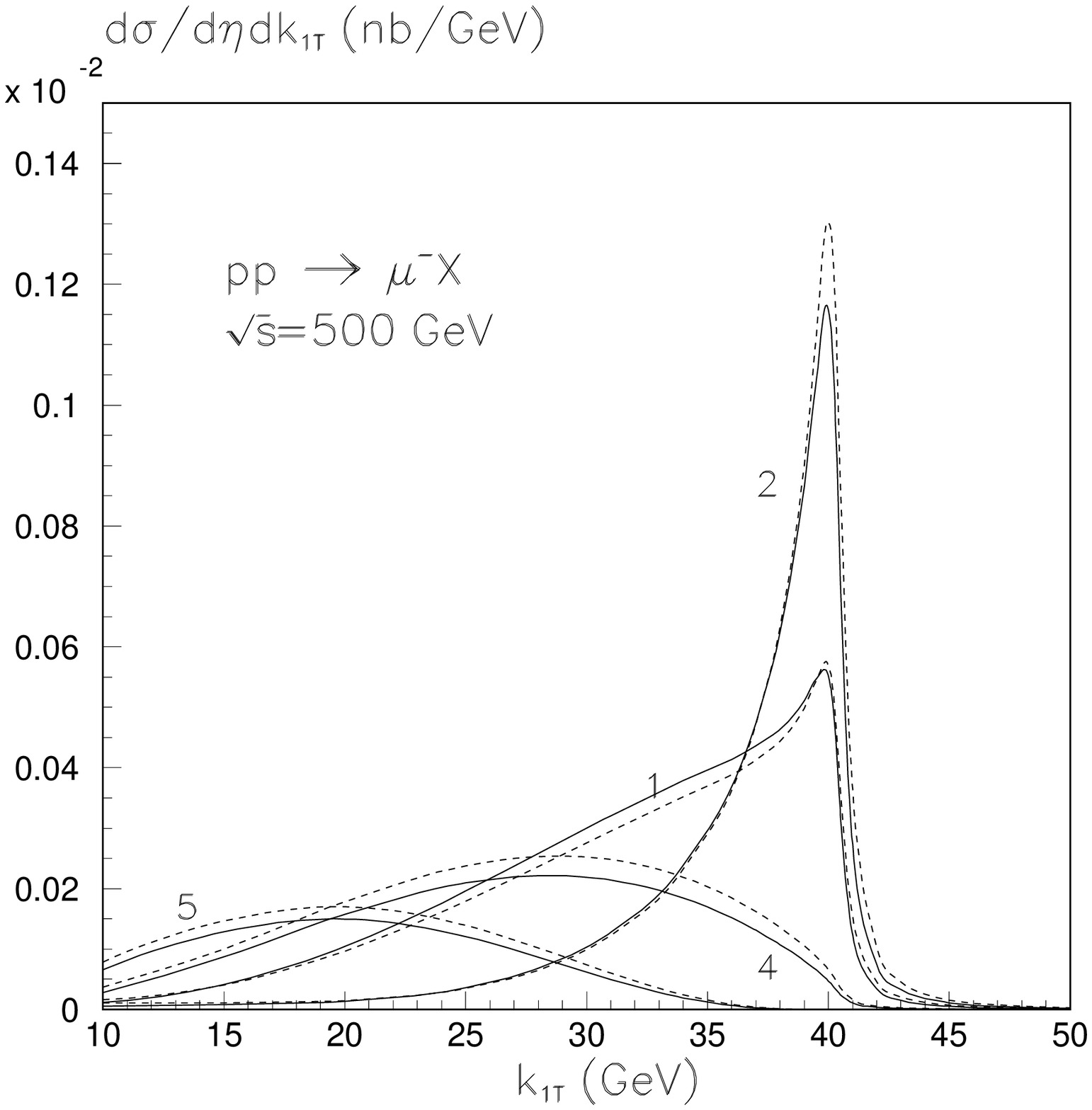} }
\end{picture}
\\[45mm]
\begin{picture}(60,60)
\put(-20,0){
\epsfxsize=6cm
\epsfysize=6cm
\epsfbox{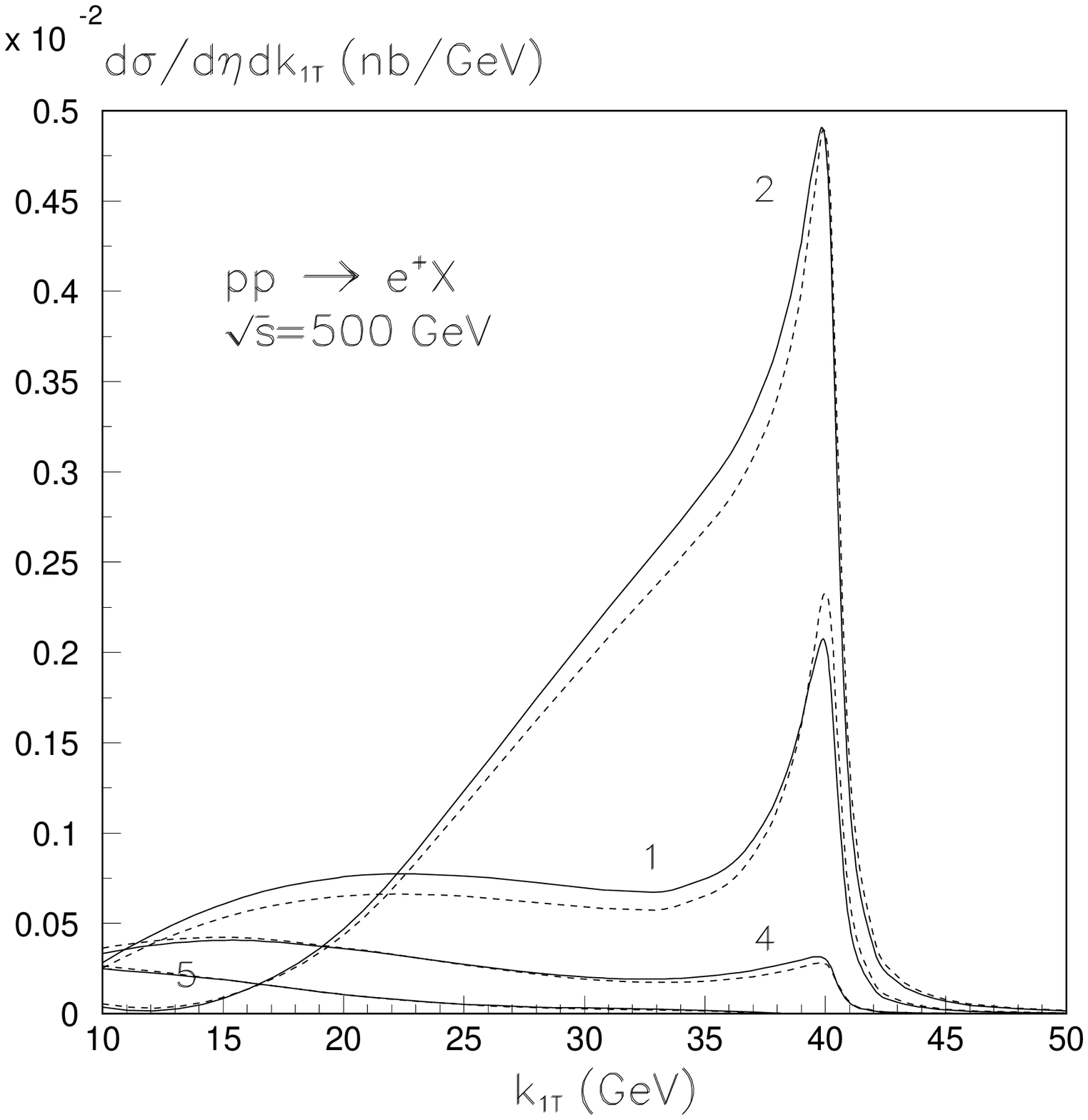} }
\end{picture}
&
\begin{picture}(60,100)
\put(110,0){
\epsfxsize=6cm
\epsfysize=6cm
\epsfbox{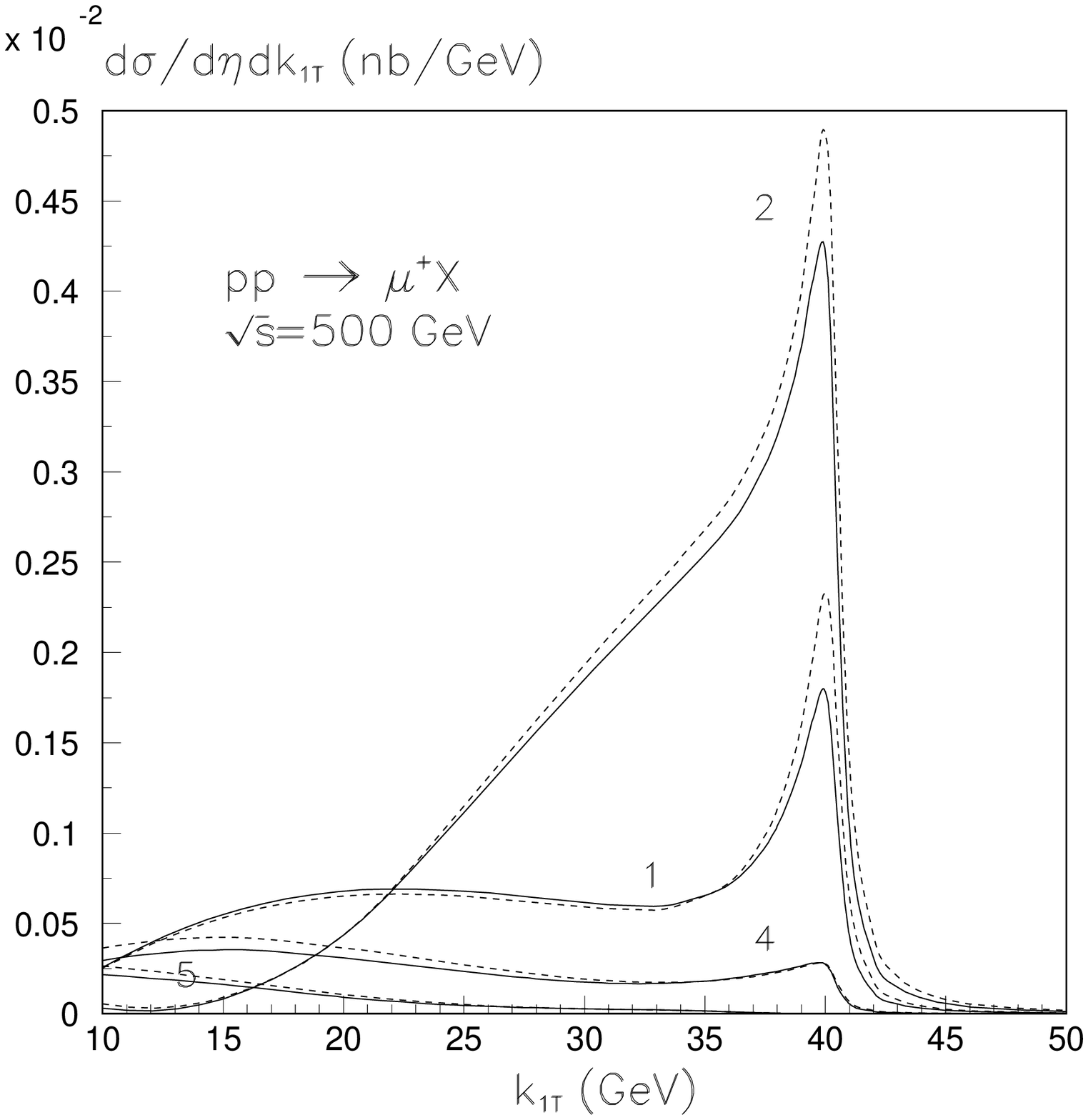} }
\end{picture}
\end{tabular}
\vspace*{1mm}
\caption{\protect\it
Spin averaged part of the double differential cross sections for
$\stackrel{\ }{p} \stackrel{\rightarrow}{p}
\rightarrow l^{\pm}X$ at $\sqrt{S}=500$ GeV, $\Delta \Phi=2\pi$
(RHIC) as a function of ${k_1}_T$ for the pseudo-rapidity
$\eta=-1$ (curve 1), $\eta=0$ (curve 2), $\eta=1.5$ (curve 4),
$\eta=2$ (curve 5) in the case of $e^{\pm}$, $\mu^{\pm}$ final states.
Shown on the figures are the cross sections in the Born approximation
(dashed lines) and taking into account the total EWC (solid lines).
We use the GRV94 proton parametrization of parton distribution
functions \cite{GRV94}.
}
\label{7}
\end{figure}


\begin{figure}
\vspace{15mm}
\begin{tabular}{cc}
\begin{picture}(60,60)
\put(-20,-40){
\epsfxsize=6cm
\epsfysize=6cm
\epsfbox{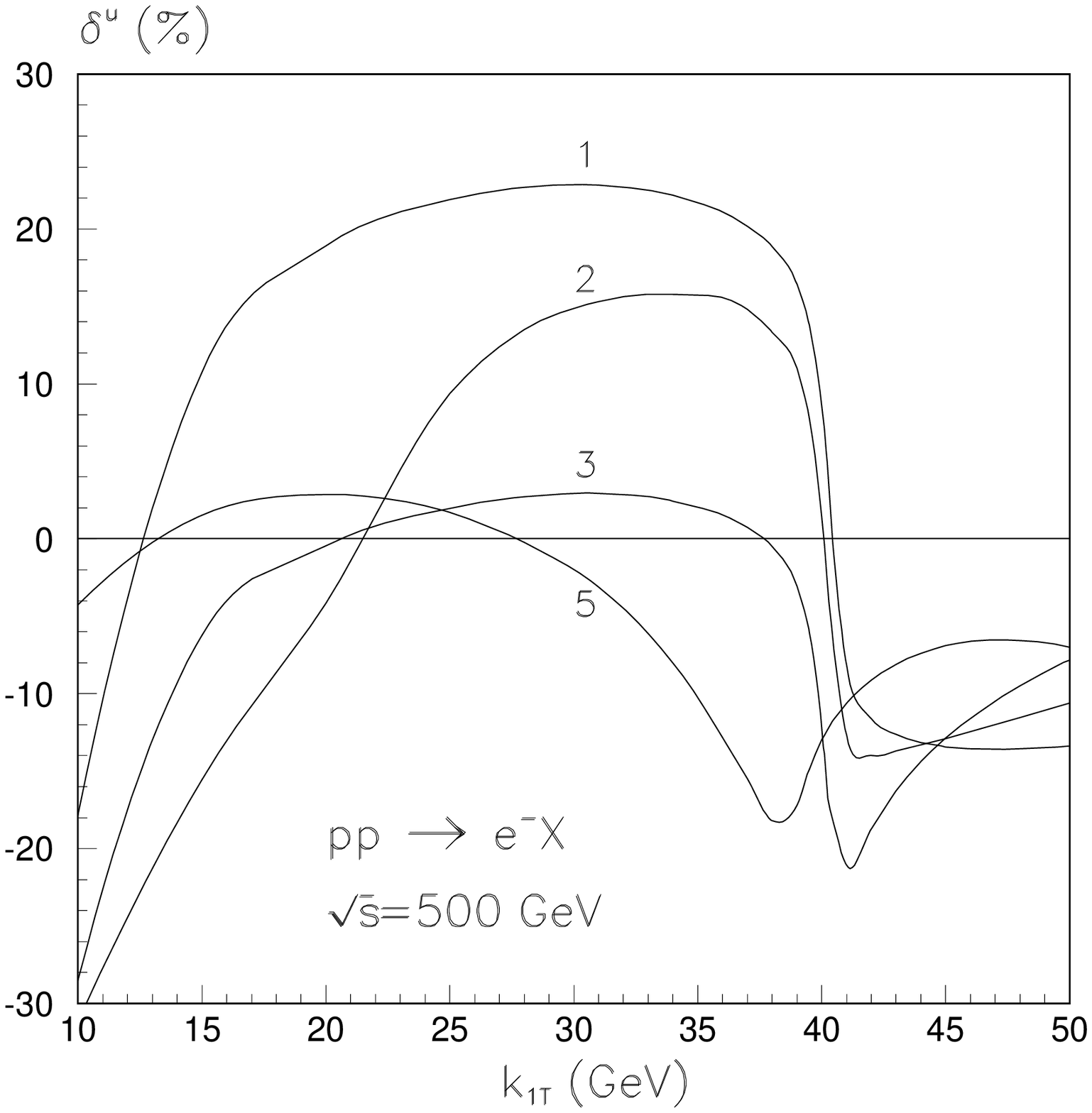} }
\end{picture}
&
\begin{picture}(60,100)
\put(110,-40){
\epsfxsize=6cm
\epsfysize=6cm
\epsfbox{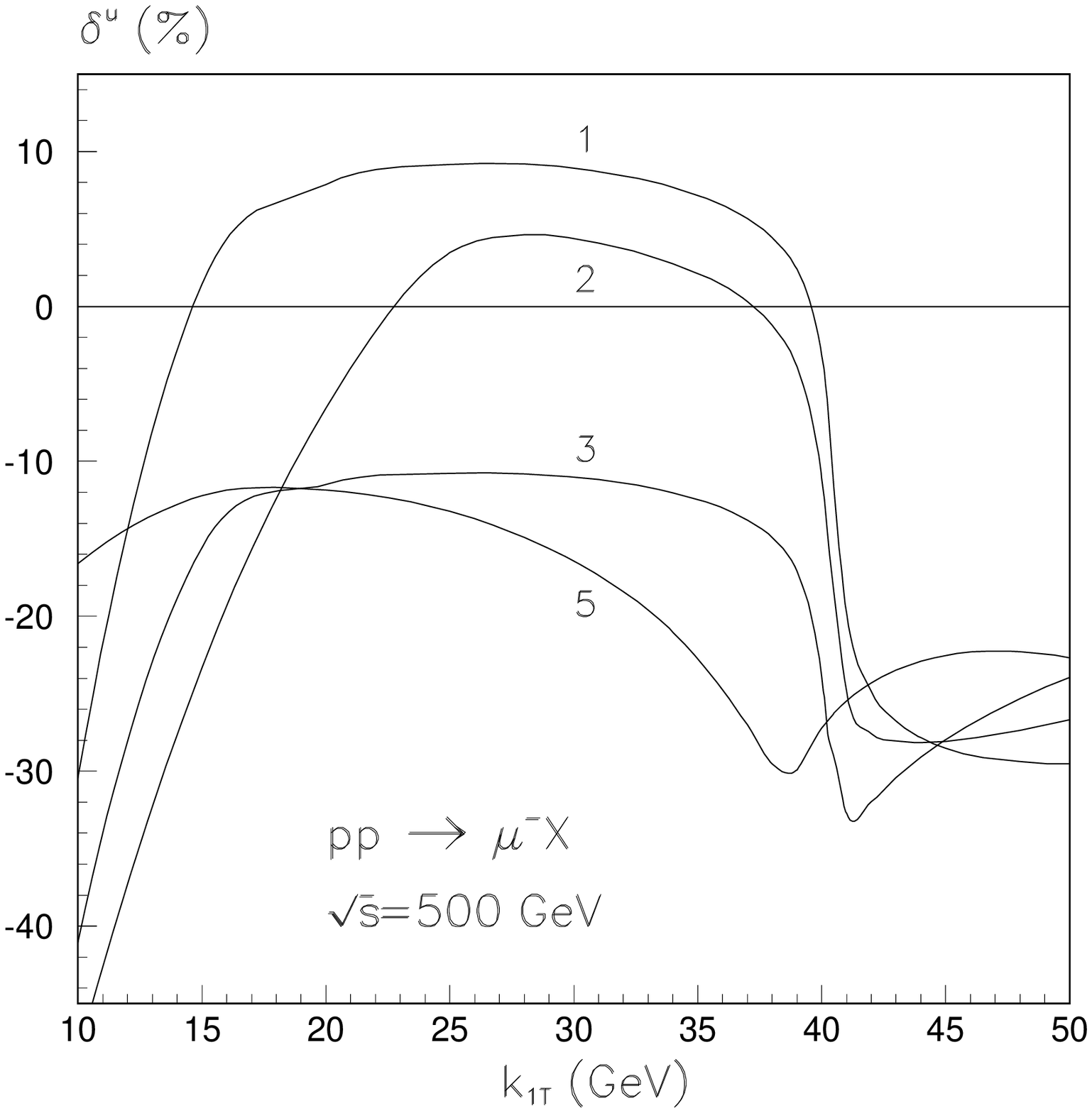} }
\end{picture}
\\[45mm]
\begin{picture}(60,60)
\put(-20,0){
\epsfxsize=6cm
\epsfysize=6cm
\epsfbox{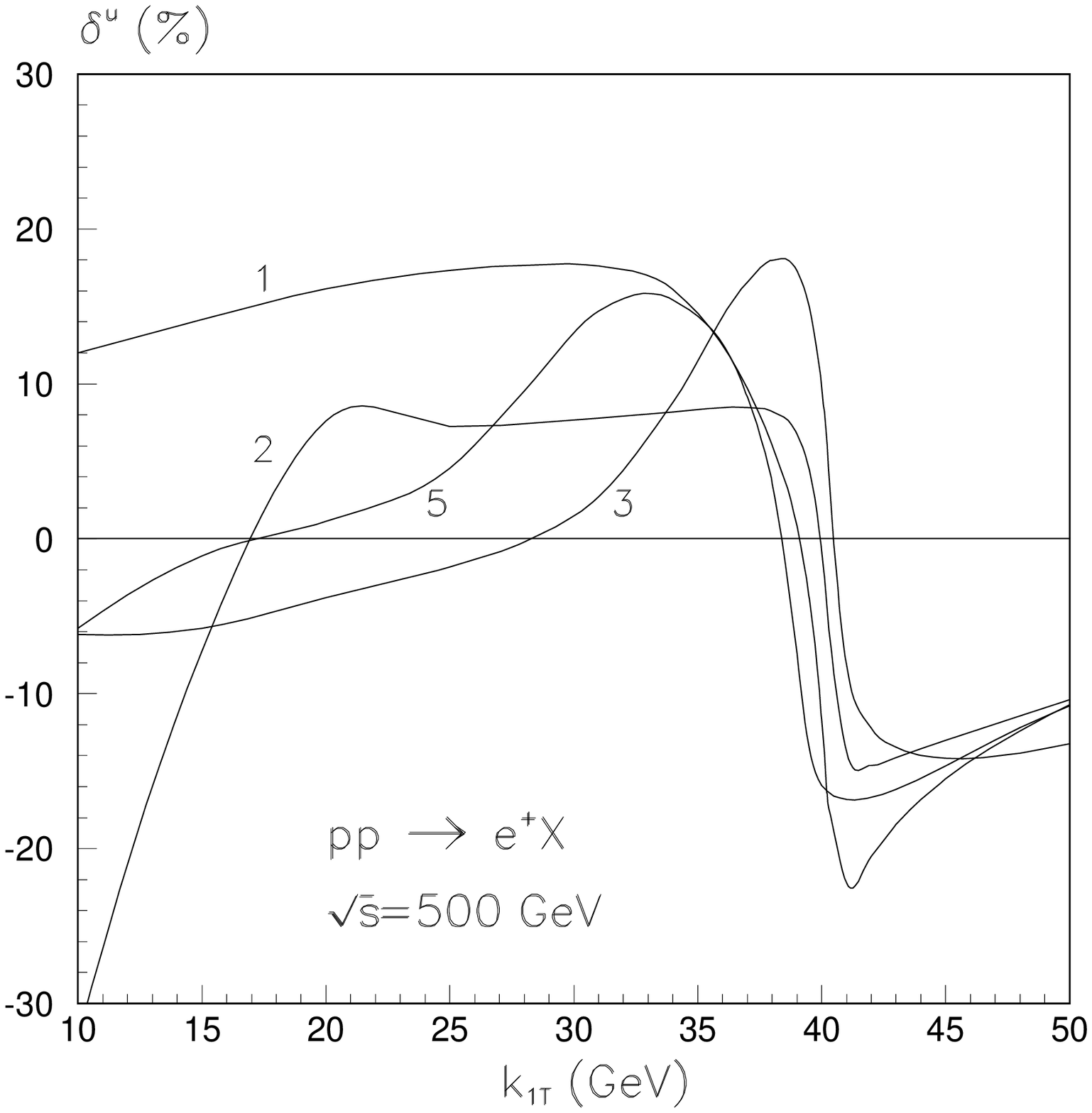} }
\end{picture}
&
\begin{picture}(60,100)
\put(110,0){
\epsfxsize=6cm
\epsfysize=6cm
\epsfbox{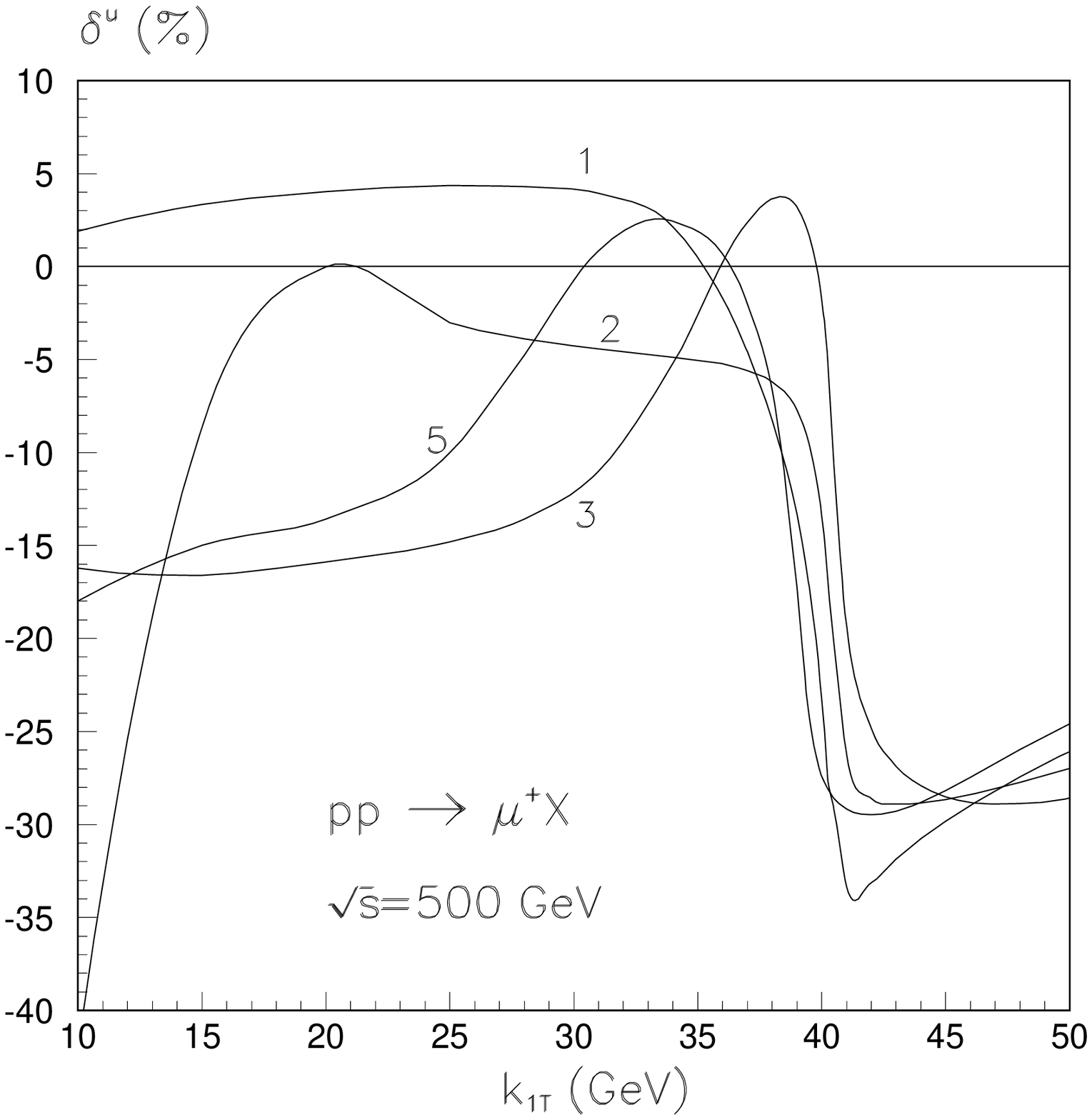} }
\end{picture}
\end{tabular}
\vspace*{15mm}
\caption{\protect\it
Corrections $\delta^u$ for
$\stackrel{\ }{p}\stackrel{\rightarrow}{p}
\rightarrow l^{\pm}X$ at RHIC kinematics as a function of ${k_1}_T$
for the pseudo-rapidity
$\eta=-1$ (curve 1), $\eta=0$ (curve 2), $\eta=1$ (curve 3),
$\eta=2$ (curve 5) in the case of $e^{\pm}$, $\mu^{\pm}$ final states.
The rest of the parameters is identical to that in Fig.7.
}
\label{8}
\end{figure}


\begin{figure}
\vspace{25mm}
\begin{tabular}{cc}
\begin{picture}(60,60)
\put(-20,-40){
\epsfxsize=6cm
\epsfysize=6cm
\epsfbox{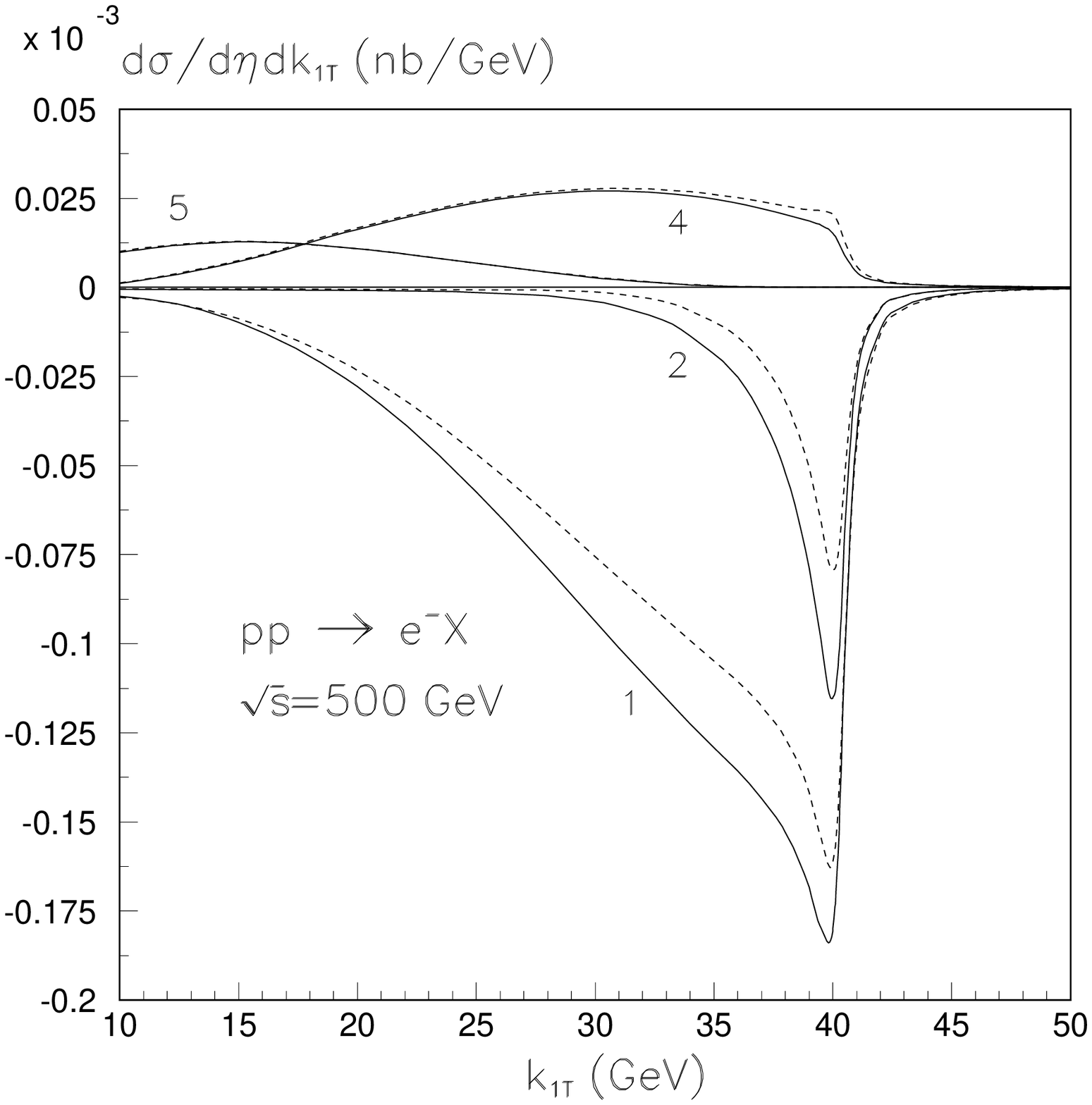} }
\end{picture}
&
\begin{picture}(60,100)
\put(110,-40){
\epsfxsize=6cm
\epsfysize=6cm
\epsfbox{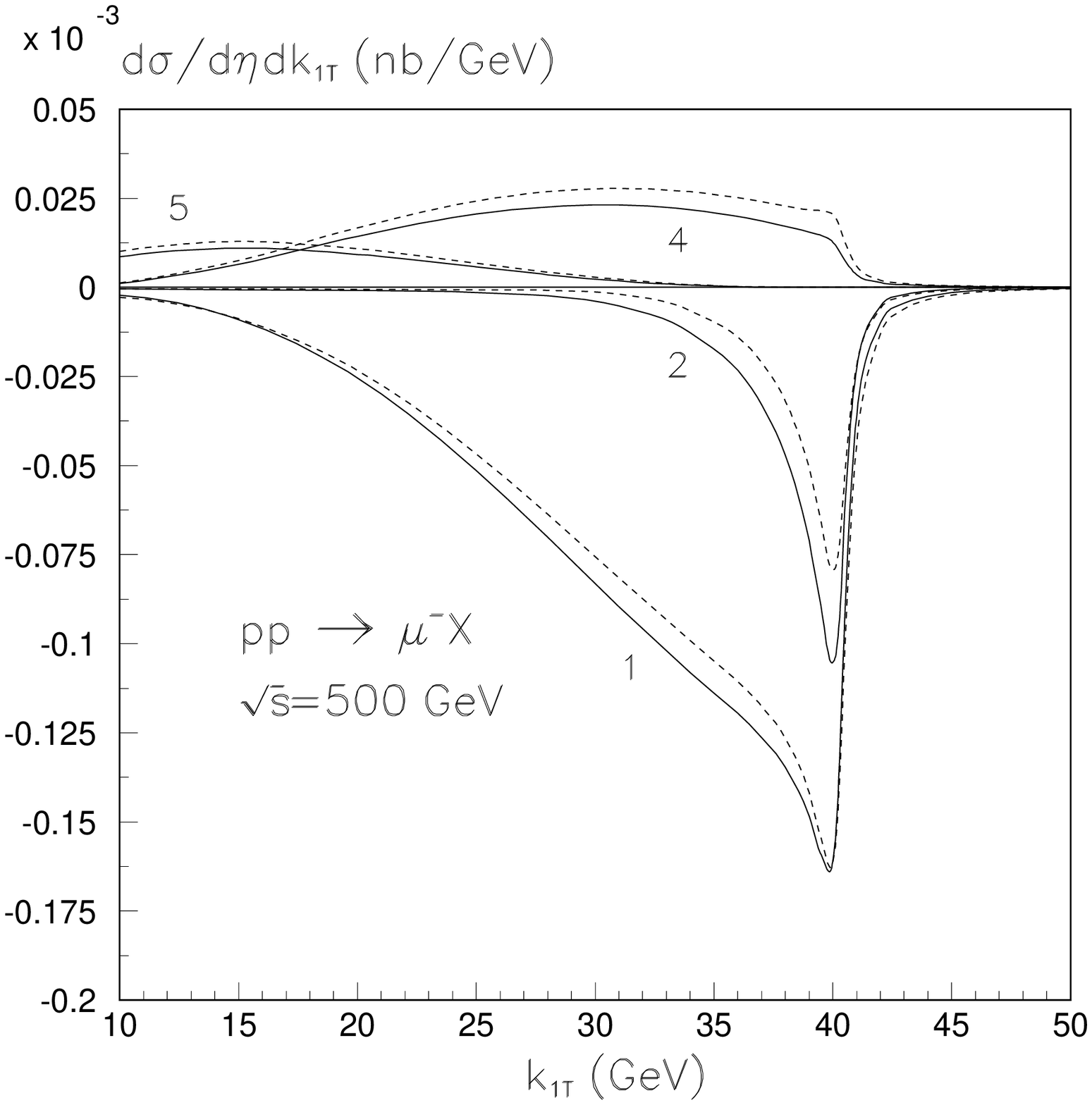} }
\end{picture}
\\[45mm]
\begin{picture}(60,60)
\put(-20,0){
\epsfxsize=6cm
\epsfysize=6cm
\epsfbox{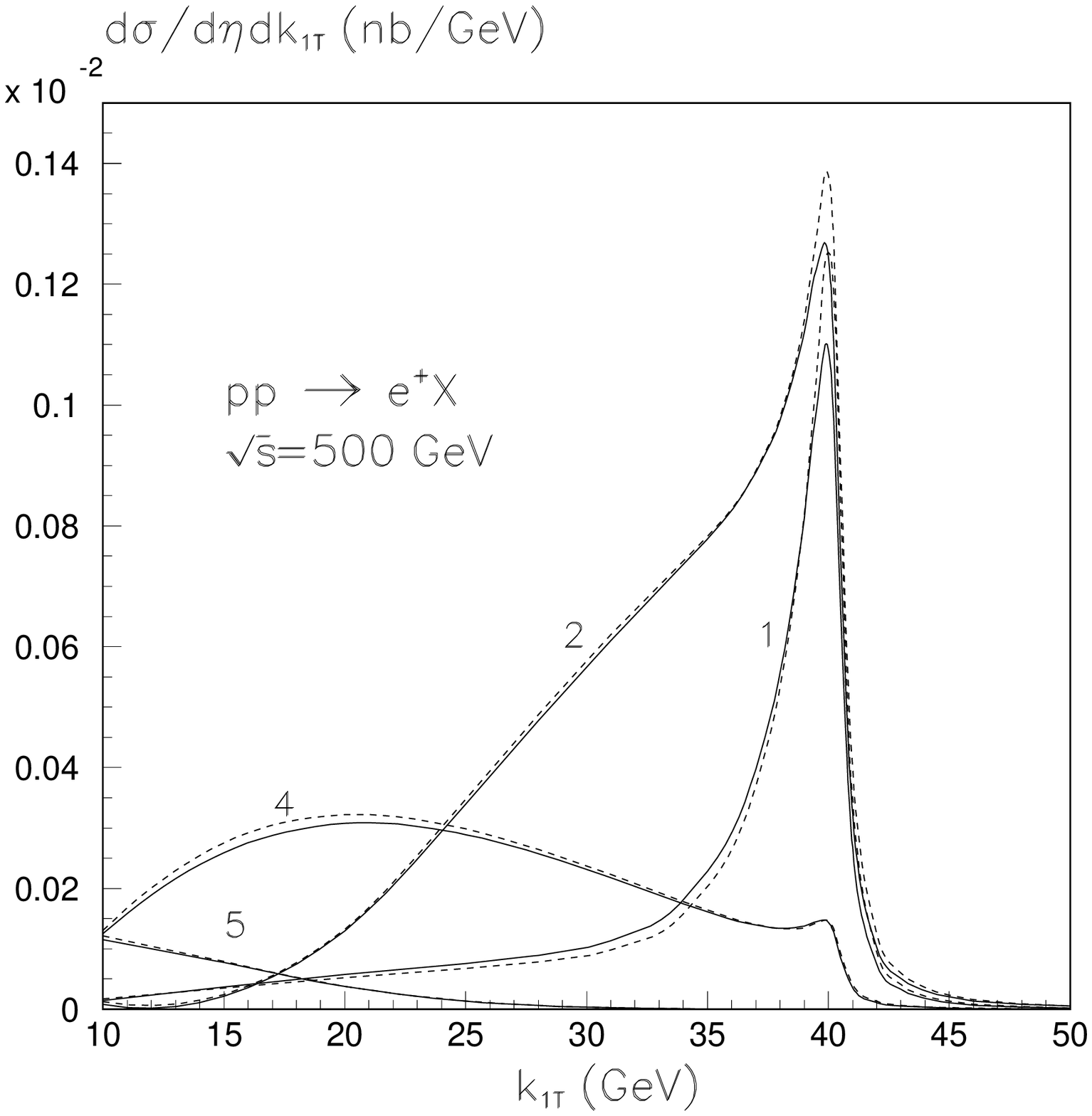} }
\end{picture}
&
\begin{picture}(60,100)
\put(110,0){
\epsfxsize=6cm
\epsfysize=6cm
\epsfbox{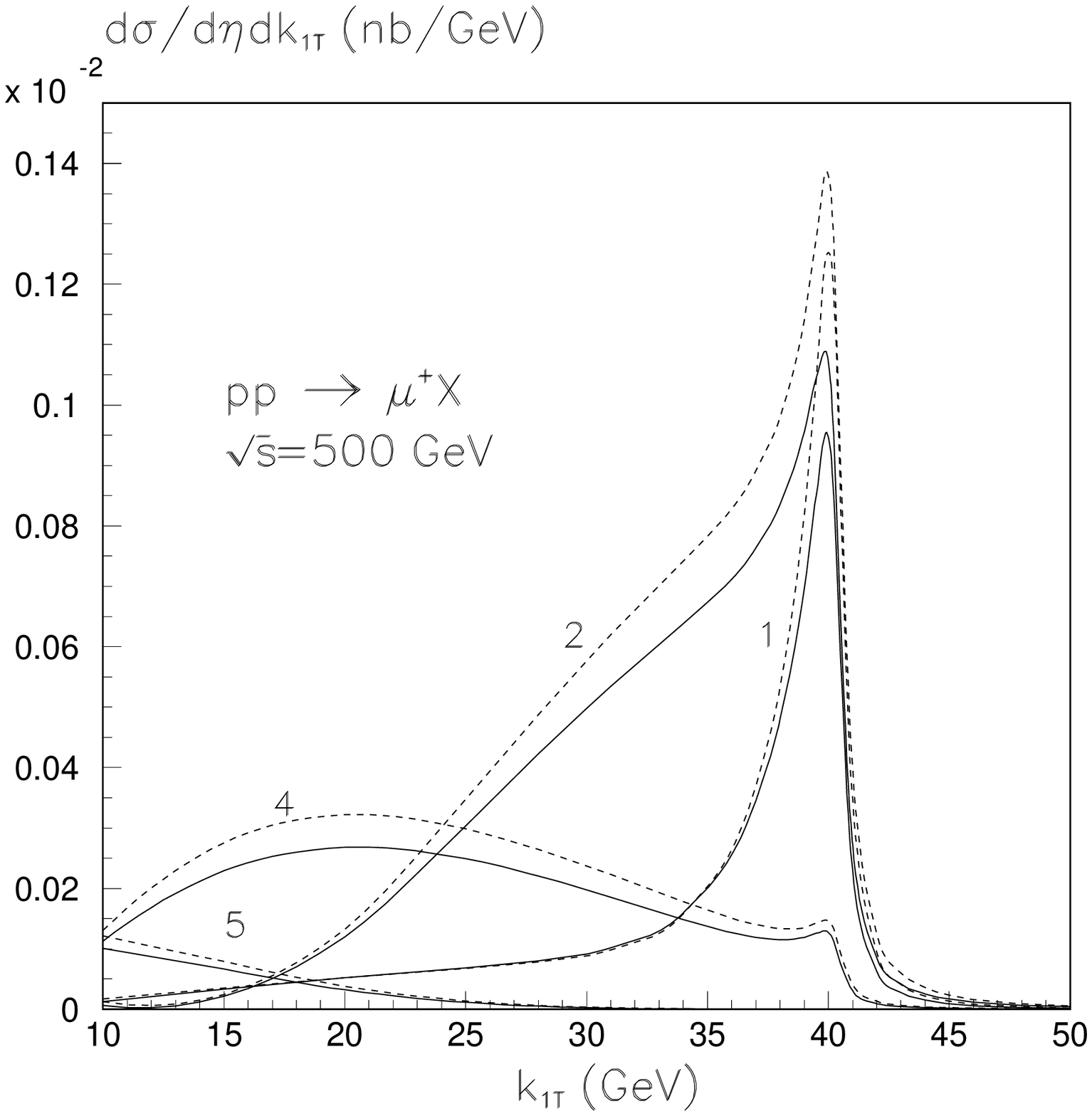} }
\end{picture}
\end{tabular}
\vspace*{5mm}
\caption{\protect\it
Polarization part of the double differential cross sections for
$\stackrel{\ }{p}\stackrel{\rightarrow}{p}
\rightarrow l^{\pm}X$ at $\sqrt{S}=500$ GeV, $\Delta \Phi=2\pi$
(RHIC) as a function of ${k_1}_T$ for the pseudo-rapidity
$\eta=-1$ (curve 1), $\eta=0$ (curve 2), $\eta=1.5$ (curve 4),
$\eta=2$ (curve 5) in the case of $e^{\pm}$, $\mu^{\pm}$ final states.
Shown on the figures are the cross sections in the Born approximation
(dashed lines) and taking into account the total EWC (solid lines).
We use the GRSV96 LO proton parametrization of polarized parton
distribution functions \cite{GRSV96}.
}
\label{9}
\end{figure}


\begin{figure}
\vspace{25mm}
\begin{tabular}{cc}
\begin{picture}(60,60)
\put(-20,-40){
\epsfxsize=6cm
\epsfysize=6cm
\epsfbox{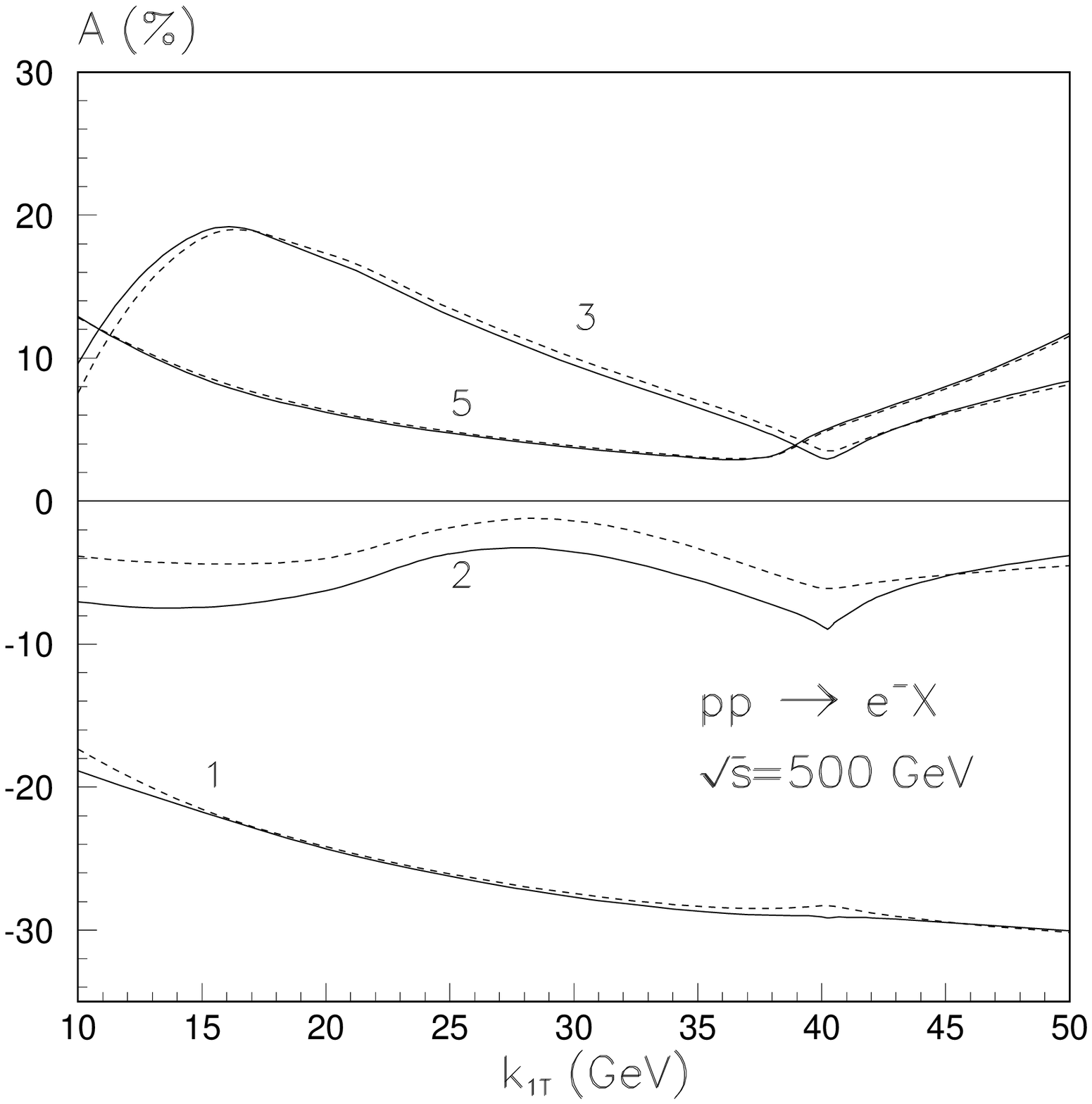} }
\end{picture}
&
\begin{picture}(60,100)
\put(110,-40){
\epsfxsize=6cm
\epsfysize=6cm
\epsfbox{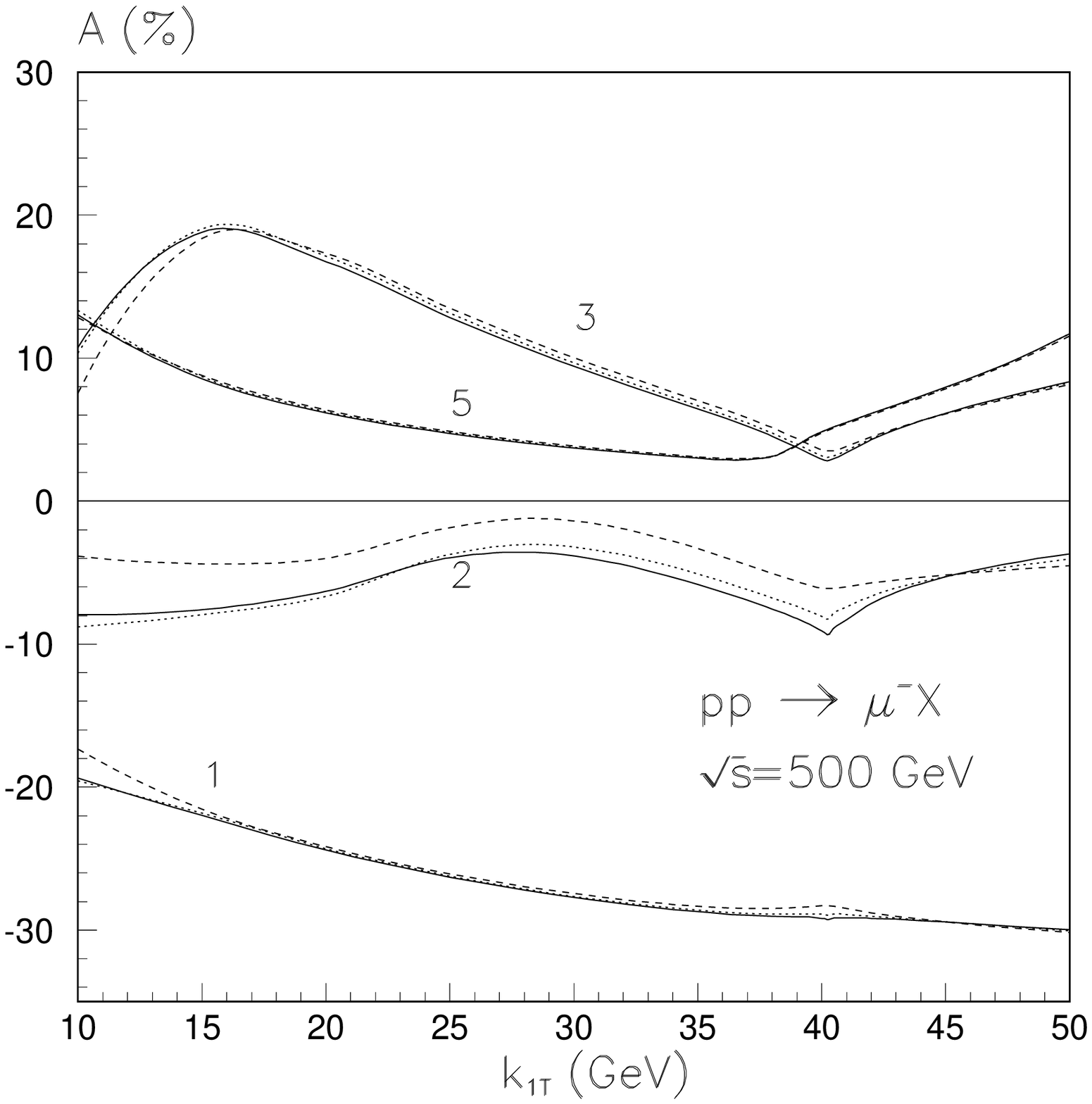} }
\end{picture}
\\[45mm]
\begin{picture}(60,60)
\put(-20,0){
\epsfxsize=6cm
\epsfysize=6cm
\epsfbox{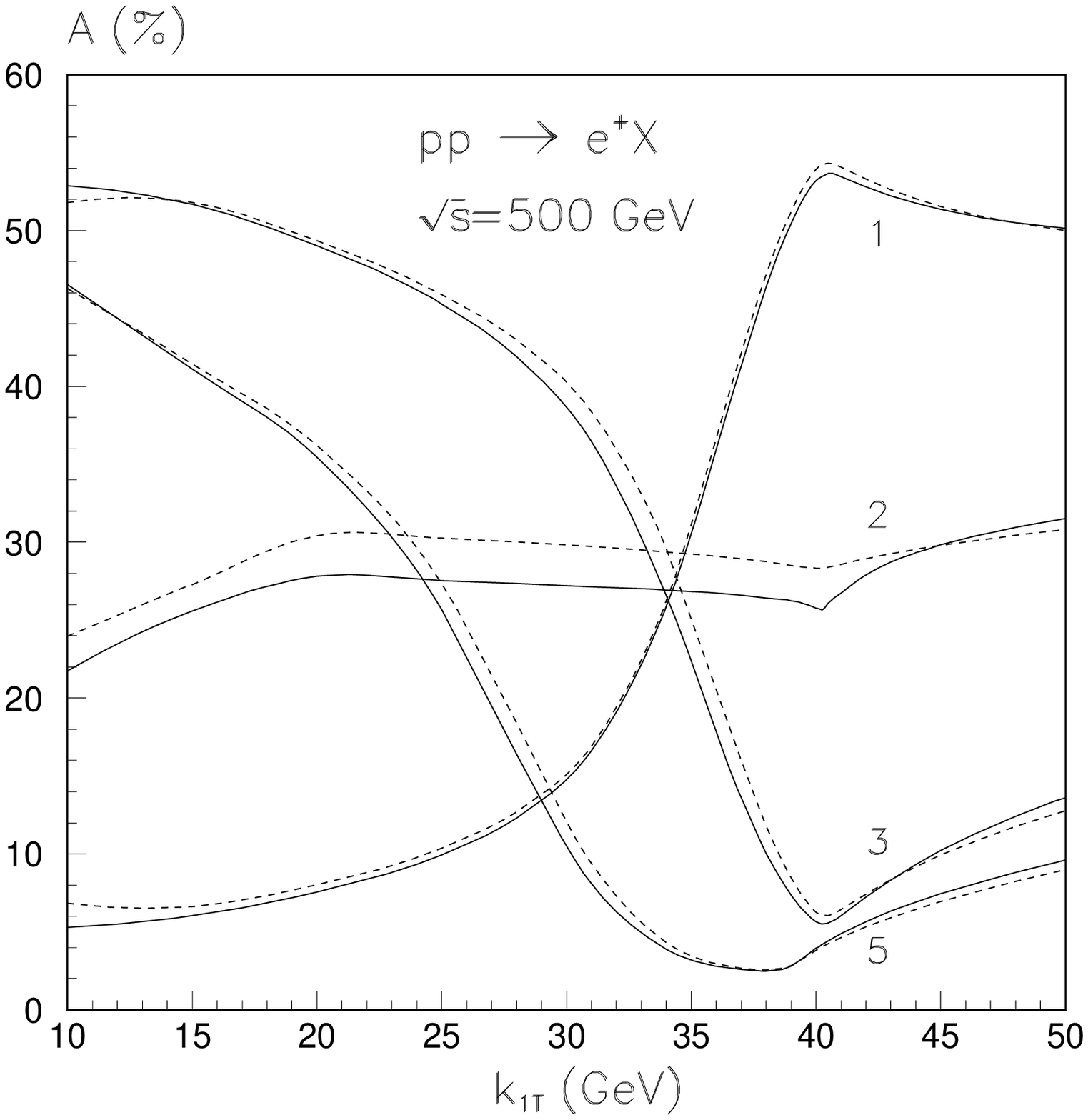} }
\end{picture}
&
\begin{picture}(60,100)
\put(110,0){
\epsfxsize=6cm
\epsfysize=6cm
\epsfbox{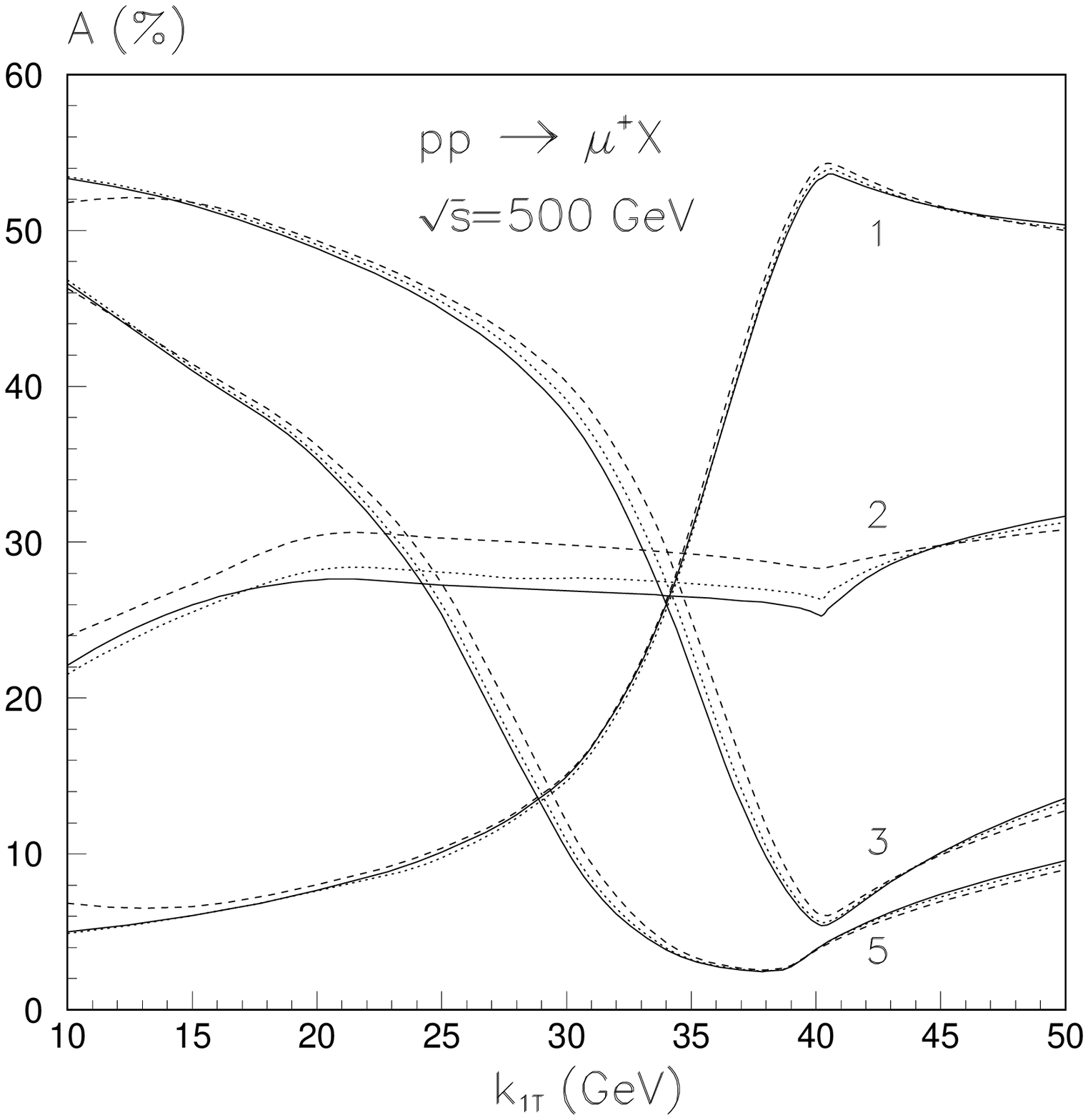} }
\end{picture}
\end{tabular}
\vspace*{5mm}
\caption{\protect\it
Single spin asymmetries for
$\stackrel{\ }{p}\stackrel{\rightarrow}{p}
\rightarrow l^{\pm}X$ at RHIC kinematics as a function of ${k_1}_T$
for the pseudo-rapidity
$\eta=-1$ (curve 1), $\eta=0$ (curve 2), $\eta=1$ (curve 3),
$\eta=2$ (curve 5)
in the Born approximation (dashed lines) and taking
into account the total EWC:
solid lines for the current quark masses,
and dotted lines for the extreme choice $m_u=m_d=0.33 GeV$.
The rest of the parameters is identical to that in Fig.9.
}
\label{10}
\end{figure}


\end{document}